\documentclass[11pt]{article}

\usepackage{authblk}
\usepackage{amsmath}
\usepackage{amsfonts}
\usepackage{amssymb}
\usepackage{mathrsfs}
\usepackage{slashed}
\usepackage{color}
\usepackage{mathtools}
\usepackage{marvosym}
\usepackage{amssymb}
\usepackage{dsfont}
\usepackage{slashed}
\usepackage{fullpage}
\usepackage{accents}
\usepackage{cite}
\usepackage[colorlinks=true,linkcolor=blue,citecolor=magenta,linktocpage=true]{hyperref}
\usepackage{titlesec}
\titleformat*{\section}{\normalsize\bfseries}
\titleformat*{\subsection}{\normalsize\bfseries}
\titleformat*{\subsubsection}{\normalsize\bfseries}
\usepackage{xcolor}

\usepackage{stackengine}
\usepackage{scalerel}
\usepackage{xcolor}
\newcommand\dangersign[1][2ex]{%
  \renewcommand\stacktype{L}%
  \scaleto{\stackon[1.3pt]{\color{red}$\triangle$}{\tiny !}}{#1}%
}

\DeclareMathAlphabet{\bbvar}{U}{BOONDOX-ds}{m}{n}
\makeatletter
\renewcommand{\@dotsep}{10000}
\makeatother

\usepackage{caption}

\newcommand{\va}{\scriptscriptstyle}
\newcommand{\la}{\label}
\newcommand{\bee}{\nopagebreak[3]\begin{equation*}}
\newcommand{\be}{\begin{equation}}
\newcommand{\ee}{\end{equation}}

\newcommand{\eee}{\end{equation*}}
\newcommand{\beq}{\begin{eqnarray}}
\newcommand{\eeq}{\end{eqnarray}}
\newcommand{\bea}{\begin{eqnarray}}
\newcommand{\eea}{\end{eqnarray}}
\newcommand{\baa}{\nopagebreak[3]\begin{eqnarray*}}
\newcommand{\eaa}{\end{eqnarray*}}

\def\be#1\ee{\begin{align}#1\end{align}}

\def\q{\qquad}
\def\f{\frac}
\def\eps{\epsilon}
\def\teps{\tilde{\epsilon}}

\def\ip{\lrcorner\,}
\def\pa{\partial}
\def\ipp{\ip}

\def\per{\text{\tiny{$\perp$}}}
\def\para{\text{\tiny{$\parallel$}}}

\def\rd{\mathrm{d}}

\def\un{\underline{n}}
\def\us{\underline{s}}

\def\vphi{\varphi}
\def\bvphi{\boldsymbol{\varphi}}
\def\bd{\boldsymbol{\delta}}

\newcommand{\R}{\mathbb{R}}
\newcommand{\RR}{\mathbb{R}}
\newcommand{\C}{\mathbb{C}}

\def\B{\mathcal{B}}
\def\BF{\mathrm{BF}}
\def\EH{\mathrm{EH}}

\def\DPS{\mathrm{DPS}}

\def\ECH{\mathrm{ECH}}
\def\GR{\mathrm{GR}}
\def\GH{\mathrm{GH}}

\def\BF{\mathrm{BF}}
\def\ts{\tilde{s}}
\def\tb{\tilde{b}}
\def\tS{\tilde{S}}
\def\tB{\tilde{B}}
\def\btE{\bar{\tilde{E}}}
\def\tbE{\tilde{\bar{E}}}
\def\tE{\tilde{E}}

\def\tK{\tilde{K}}
\def\tP{\tilde{P}}

\def\bTh{\boldsymbol\Theta}

\def\CA{\mathcal{A}}
\def\CB{\mathcal{B}}
\def\CC{\mathcal{C}}

\def\CH{\mathcal{H}}

\def\CL{\mathcal{L}}

\def\CQ{\mathcal{Q}}
\def\CR{\mathcal{R}}

\def\NN{\mathsf{N}}

\def\n{\mathsf{n}}

\def\JJ{\mathsf{J}}
\def\ext{\mathrm{ext}}

\def\SU{\text{SU}}

\def\SL{\text{SL}}
\def\su{\mathfrak{su}}

\def\sll{\mathfrak{sl}}

\def\e{\mathsf{e}}

\def\te{\tilde{e}}
\def\bte{\bar{\tilde{e}}}

\def\tbe{\tilde{\bar{e}}}

\def\BB{\mathsf{B}}

\numberwithin{equation}{section}
 
\begin{document}

\title{\Large{\textbf{\sffamily Edge modes of gravity - II:\\Corner metric and Lorentz charges}}}
\author{\sffamily Laurent Freidel$^1$, Marc Geiller$^2$, Daniele Pranzetti$^1$}
\date{\small{\textit{
$^1$Perimeter Institute for Theoretical Physics,\\ 31 Caroline Street North, Waterloo, Ontario, Canada N2L 2Y5\\
$^2$Univ Lyon, ENS de Lyon, Univ Claude Bernard Lyon 1,\\ CNRS, Laboratoire de Physique, UMR 5672, F-69342 Lyon, France\\}}}

\maketitle

\begin{abstract}
In this second paper of the series we continue to spell out a new program for quantum gravity, grounded in the notion of corner symmetry algebra and its representations. Here we focus on tetrad gravity and its corner symplectic potential. We start by performing a detailed decomposition of the various geometrical quantities appearing in BF theory and tetrad gravity. This provides a new decomposition of the symplectic potential of BF theory and  the simplicity constraints. We then show that the dynamical variables of the tetrad gravity corner phase space are  the internal normal to the spacetime foliation, which is conjugated to the boost generator, and the corner coframe field. This allows us to derive several key results. First, we construct the corner Lorentz charges. In addition to sphere diffeomorphisms, common to all formulations of gravity, these charges add a local $\mathfrak{sl}(2,\mathbb{C})$ component to the corner symmetry algebra of tetrad gravity. Second, we also reveal that the corner metric satisfies a local $\mathfrak{sl}(2,\mathbb{R})$ algebra, whose Casimir corresponds to the corner area element. Due to the space-like nature of the corner metric, this Casimir belongs to the unitary discrete series, and its spectrum is therefore quantized. This result, which reconciles discreteness of the area spectrum with Lorentz invariance, is proven in the continuum and without resorting to a bulk connection. Third, we show that the corner phase space explains why the simplicity constraints become non-commutative on the corner. This fact requires a reconciliation between the bulk and corner symplectic structures, already in the classical continuum theory. Understanding this leads inevitably to the introduction of edge modes.
%

\end{abstract}

\thispagestyle{empty}
\newpage
\setcounter{page}{1}

\hrule
\tableofcontents
\vspace{0.7cm}
\hrule

\newpage

\section{Introduction}
\la{sec:1}

We have recently proposed in the companion paper \cite{Freidel:2020xyx} a new roadmap towards quantum gravity, grounded in the notion of representation of the corner symmetry algebra $\mathfrak{g}_S$. This is a universal notion of symmetry algebra at codimension-2 corners of any subregion of space, which exists independently of boundary conditions. It can be defined at the classical level, and  demanding that the corner symmetries survive quantization gives a new organizing principle for understanding quantum gravity. 
 
 \subsection{Motivations}

Our emphasis on the corner algebra is motivated by the lessons and deep insights coming from two radically different approaches to quantum gravity, namely the teachings of AdS/CFT holography and of loop quantum gravity (LQG). This new perspective, which we call \textit{local holography},\footnote{The term local holography has also been used in \cite{Dittrich:2017hnl,Dittrich:2017rvb,Dittrich:2018xuk}. We use it here in a related but different manner.} can be seen as a merging of some of the key concepts of both approaches. It can also be seen as a new beginning, with an entirely different conceptual perspective and new technical tools, allowing to address some critical shortcomings of both approaches and to reconcile their principles and objectives.
 
In AdS/CFT, one postulates that the bulk information can be encoded in terms of observables living on an asymptotic boundary with  specific dynamics. The challenge is then to reconstruct the quantum bulk geometry, or even produce a consistent definition of what this could be \cite{deBoer:1999tgo, deHaro:2000vlm,  Maldacena:2003nj,Marolf:2004fy,Hamilton:2005ju,Hamilton:2006az,Marolf:2008mf,Comp_re_2008,DeJonckheere:2017qkk,Dong:2016eik}. This question is still open despite interesting recent advances in lower-dimensional gravity \cite{Freidel:2008sh, McGough:2016lol, Tolley:2019nmm, Mazenc:2019cfg, Saad:2019lba, Stanford:2019vob, Almheiri:2019qdq, Mertens:2020hbs, Iliesiu:2020zld, Maxfield:2020ale,Witten:2020wvy}. What is crucially missing in all these descriptions is an understanding of the nature of micro-states of quantum geometry. In local holography, we propose to decompose the bulk of spacetime into a collection of subregions, and to attach a symmetry algebra to the corner of each subregion. The corner symmetry charges encode a coarse-graining of the information inside each region it encloses. The corner Hilbert space forms an irreducible representation of the local corner symmetry algebra, and choices of states in this corner Hilbert space then encode quantum geometries. One of the  goals of this program is to show the conjectural claim that these \textit{continuous} and \textit{covariant} corner Hilbert spaces can be taken to be finite-dimensional, with a size depending on the value of the Casimirs of the corner algebra and the quasi-local energy. In this case, the coarse-graining would no longer be an approximation, and can become exact at the quantum level.

One  also expects to implement the bulk constraints as conservation laws for the local corner charges. At the quantum level, these conservation laws are conjectured to be understood in terms of generalized intertwiners, which defines a continuous version of the fusion product for corner Hilbert spaces. This gives a concrete implementation of the profound albeit rather vague statement that one can get ``spacetime from entanglement of quantum information'' \cite{VanRaamsdonk:2010pw,Almheiri:2014lwa,Pastawski:2015qua,VanRaamsdonk:2016exw,swingle2018,Dong:2018cuv,Almheiri:2018xdw,VanRaamsdonk:2018zws}. The program of constraint reconstruction as corner charge conservation was initiated in \cite{Freidel:2015gpa,Freidel:2016bxd,Freidel:2018pvm,Freidel:2019ees} with a focus on the kinematical constraints. A similar point of view is also adopted in the group field theory approach \cite{Oriti:2014uga}, and some key ideas relating dynamics and fusion product have been worked out in \cite{Dittrich:2013xwa,Delcamp:2016yix,Dittrich:2016typ,Cunningham:2020uco}. The two central points of the program which we propose are still conjectural and should be thought of at this point as directions to explore. In this series of papers we focus on some key kinematical and semi-classical aspects as the first steps in this direction.
 
The idea that quantum gravity should assign to every local region a local Hilbert space labeling state of quantum geometry associated with the boundary surface is not new. It is one of the central and most crucial aspects of LQG \cite{Thiemann:2001yy,Ashtekar:2004eh}, as first revealed by Smolin and Krasnov \cite{Smolin:1995vq,Krasnov:1996tb}, and successfully used in the black hole micro-states counting \cite{Ashtekar:1997yu,Domagala:2004jt,Meissner:2004ju,Agullo:2008yv,Engle:2011vf}. It also the central theme of tensor network realizations of spacetime as a quantum process \cite{Dittrich:2013xwa,Donnelly:2016qqt, May_2017,Qi:2018shh, Bao:2018pvs, Chirco:2019dlx} and it has made its appearance in attempts to extend AdS/CFT to finite boundaries \cite{Miyaji:2015yva,Takayanagi:2019tvn}. In LQG, one postulates that quantum geometry is supported on a network of distributional configurations (loops) and that the corresponding states of quantum geometry are spin network states carrying $\SU(2)$ representations living on the links of the network. When intersecting this network with a sphere one obtains a collection of punctures labeling representations, which defines a Hilbert space attached to the sphere (the intertwiner space). This Hilbert space, which carries representations of a product of $\SU(2)$'s, is a particular and simple example of the discretization of a corner Hilbert space \cite{Edge-Mode-III}.
 
There are several challenges however with this approach. First, it relies on a connection formulation using the $\su(2)$ Ashtekar--Barbero connection \cite{Barbero:1994ap, Barbero:1994an}, which exists only in the tetrad formalism and in the so-called time gauge. This fact makes it almost impossible  to transpose the results obtained there to metric gravity, where such a connection does simply not exist. Moreover, attempts to relax the time gauge and extend this connection to a manifestly Lorentz-invariant setting have at best led to ambiguous results \cite{Alexandrov:2010un,Alexandrov:2011ab}. More problematic is the reliance of this formulation on the presence of an underlying network along which one integrates the connection. The presence of this network introduces discontinuities and singularities,\footnote{There exists  a reformulation of LQG working with discrete but non-singular representations \cite{Bahr:2015bra,Dittrich:2014wda,Dittrich:2014wpa}.} which are usually rationalized as being quantum geometry effects with no classical analog \cite{Rovelli:1987df,Rovelli:1989za,Ashtekar:1991kc,Ashtekar:1992tm,Ashtekar:1994mh,Rovelli:1995ac}. Moreover, the non-commutativity of the geometrical observables, which is central to the program, can be traced back to these singularities \cite{Ashtekar:1998ak}.

One of our goals is to keep the  deep insights of  LQG, while freeing ourselves from its most unpalatable aspects, such as the built-in discretization. In particular, in the present paper we show that it is possible to have a non-commutative symmetry algebra without introducing discrete graph structures, without relying on a connection formulation, and without having to fix the time gauge. This, in turn, allows us to construct a new geometrical operator, which is the \textit{corner metric}. This then allows us to unambiguously reconcile Lorentz invariance with discreteness of area.

\subsection{Questions}\la{sec:questions}

Since the central focus of our program is the corner algebra, we first need to understand what are the boundary symmetry algebras associated with different formulations of gravity. As explained in \cite{Freidel:2020xyx}, this is achieved by decomposing the symplectic potential of each formulation of gravity into a bulk piece and a corner piece. The bulk piece is common to all formulations and defined by a momentum density conjugated to the induced metric (or induced coframe) on the slice. This yields the component $\mathrm{diff}(S)$ of the corner symmetry algebra of gravity $\mathfrak{g}_S$, corresponding to diffeomorphisms generated by tangent vector fields non-vanishing on the corner $S$. This part of the algebra is non-trivially represented in all formulations of gravity. What differentiates various formulations is then the corner symplectic structure. We reveal that different formulations carry a different set of corner charges, which provide a non-trivial representation for different  components of the corner symmetry algebra. This framework enables us to phrase the first central questions we address in this paper:

\begin{itemize}
\item[$(i)$] Can classically equivalent formulations of gravity lead to inequivalent quantizations? Is there a fundamental difference in that respect between the second order metric formulation and the first order formulation in terms of tetrads?
\end{itemize}

In \cite{Freidel:2020xyx} we have concentrated on the case of metric Einstein--Hilbert (EH) gravity and established the results summarized in the first two lines of Table \ref{table} below. Here we switch to tetrad gravity and consider the Einstein--Cartan--Holst (ECH) formulation, where Holst denotes the inclusion of a topological term whose coupling is the Barbero--Immirzi parameter $\gamma=\beta^{-1}$. The goal of the present paper is to establish the last three lines in Table \ref{table}. In order to treat the ECH case, our strategy consists of first analyzing BF theory \cite{Baez_1996} and then extanding our results to tetrad gravity by using the so-called \textit{simplicity constraints} \cite{Plebanski:1977zz,Capovilla:1991kx,Capovilla:1991qb,Reisenberger:1996ib,DePietri:1998hnx,Capovilla:2001zi}. These constraints express the $B$ field as a wedge product of the gravitational coframe field according to
\be 
B^{IJ}=(*+\beta)(e\wedge e)^{IJ}\eqqcolon E^{IJ}[e].
\ee
In the Hamiltonian analysis, these primary constraints form a second class pair with their conjugated secondary constraints \cite{Ashtekar:1991hf,Romano_1993,Peld_n_1994,BarroseSa:2000vx,Alexandrov:2000jw,Alexandrov:2002br,Alexandrov:2006wt,Cianfrani:2008zv,Alexandrov:2010un,Alexandrov:2011ab,Geiller:2011bh,Montesinos_2019}. If we don't impose  the time gauge, one finds that the construction of a Lorentz covariant connection is ambiguous \cite{Alexandrov:2001wt,Alexandrov:2010un,Alexandrov:2011ab}. This ambiguity is directly related to the question of whether discreteness of the area spectrum survives the relaxation of the time gauge \cite{Alexandrov:2001pa,Alexandrov:2002br,Geiller:2011bh}. Moreover, the difference of treatment of the simplicity constraints in canonical LQG and in the covariant spin foam approach is the source of another set of ambiguities and puzzles \cite{Alexandrov:2008da,Anza:2014tea,Dupuis:2010iq,Geiller:2011aa,Livine:2007ya,Dittrich:2010ey}. This therefore raises the second set of fundamental questions which we want to address:

\begin{itemize}
\item[$(ii)$] Is it possible to reconcile Lorentz invariance with the discreteness of the area spectrum? How can the simplicity constraints be properly imposed at the quantum level?
\end{itemize}

It is important to acknowledge that the first question is relevant in \textit{any} approach to quantum gravity. Even in a metric formulation where the issue of dealing with the simplicity constraints is bypassed, the derivation of a discrete area spectrum from the continuum theory is a central open issue, which is key in order to explain the origin of a finite entropy \cite{Bekenstein:1973ur,Ryu:2006bv}. In the context of tetrad gravity, we claim that the answer to both questions in $(ii)$ has been within reach all along. In order to realize this, one needs to shift the emphasis from the bulk to the corner. Instead of using the Holst term and the Barbero--Immirzi parameter in order to build a connection formulation in the bulk, one can push this contribution to the corner and obtain a corner phase space structure with a non-commutative coframe field. At the level of the symplectic potential, this follows essentially from the identity \cite{Thiemann:2001yy,Liu:2009em,Engle:2010kt}
\be
\beta(e\wedge e)_{IJ}\wedge\delta\omega^{IJ}\simeq-\beta\rd(e_I\wedge\delta e^I),
\ee
where the symbol $\simeq$ denotes the use of the torsion-free condition. Acceptance of this technically simple fact is forced upon us by the local holographic framework which we are developing, and it has deep conceptual implications. In particular, it implies that the connection is no longer the fundamental object required to build the kinematical Hilbert space for quantum geometry. Instead, this role is now played by the non-commutative corner coframe. This paradigm shift dissolves immediately ambiguities related to the choice of bulk connection, and, as we will explain, it enables us to also treat unambiguously the simplicity constraints (which become second class with themselves on the corner).  At the same time, it naturally explains from a continuum and semi-classical\footnote{The use of the term ``semi-classical'' in our context has to be understood in relation to the Kirillov orbit method \cite{Kirillov}, as explained in \cite{Freidel:2020xyx}.} perspective how the discreteness of the area spectrum follows from the symmetry algebra of the corner charges.

At this point, one could be worried about yet another fundamental puzzle: The fundamental geometrical fields seem to behave differently in the bulk and at the corner. More precisely, while the fluxes classically Poisson commute in the bulk, they become non-commutative at the corner after imposing the Gauss law \cite{Cattaneo:2016zsq}. In fact this is not a new concern, and it was pointed out for the flux observables already early on in the LQG framework \cite{Ashtekar:1998ak}. There, it was believed that the resolution of the tension lies entirely in the fact that the singular loopy excitations of quantum geometry develop quantum features (in this case non-commutativity) which have no classical analog \cite{Rovelli:1994ge,Ashtekar:1996eg}. If we want to let go of this gap between classical and quantum states, the challenge is then to find a consistent explanation for this puzzle already at the semi-classical and continuum level. If we believe that the corner coframe field plays a fundamental role in describing the gravitational degrees of freedom both at the classical and at the quantum level, then we are forced to face the third central question addressed in this paper:

\begin{itemize}
\item[$(iii)$] How can we reconcile the different bulk and corner canonical Poisson structures with the continuity of the coframe field?
\end{itemize}

The answer to this question will be grounded in the notion of edge modes. In order to get there, let us now summarize the various results which we obtain, and how they build upon each other in order to arrive at the edge modes.

\subsection{Summary of the results}

We now summarize three of our main results, which as we will explain encode essentially the answer to the three questions raised above.

Our study of the classical phase space of ECH gravity is based on the covariant phase space formalism. In this approach the discussion about the treatment of the second class constraints of the Hamiltonian theory in the bulk is bypassed because the formalism is simply on-shell. Still, one has to find a proper decomposition of the phase space variables which enables us to access the equations of motion and to decompose the symplectic potential. 
As the bulk + corner decomposition of the symplectic potential for tetrad gravity is a technically more involved and subtle process than in the metric case, with more constraints involved, we proceed step by step.

We start with BF theory and perform a decomposition into tangential/normal and horizontal/vertical components of all the geometrical objects defining the theory. This leads us to the notion of \textit{boost} and \textit{spin BF coframes}, which in turn provides a decomposition of the BF symplectic potential into a bulk term and a corner term. This is the first time such a bulk + corner decomposition of BF theory is performed. In particular, this reveals that BF theory possesses an additional canonical pair in the bulk as compared to gravity, which turns out to vanish on-shell when imposing the bulk simplicity constraints turning BF theory into ECH tetrad gravity. The novelty of our treatment of ECH gravity is that it keeps the Holst contribution on the corner and parametrizes the bulk degrees of freedom unambiguously in terms of the universal GR symplectic potential \cite{Freidel:2020xyx}. When relaxing the requirement of the time gauge, the internal normal is part of the corner phase space, as well as a contribution from the Holst term. This is expressed by the first main result of the paper, derived in Section \ref{sec:ECH-potential}, which is
\be\la{res1}
(i)\q\Omega_\ECH\simeq\int_\Sigma\delta\tP_I\wedge\delta\te^I+\int_S\left(\delta \tE_I\wedge\delta n^I-\f{\beta}{2}\delta\te_I\wedge\delta\te^I\right).
\ee
Elements of this decomposition, including the central role of the internal normal on the corner phase space, have already appeared in the literature 
\cite{Peld_n_1994, 0264-9381-4-5-011,Thiemann:2001yy, Liu:2009em, Engle:2010kt,bianchi2012horizon,Bodendorfer:2013hla, Bodendorfer_2013,Bodendorfer:2013jba, Wieland_2014,Wieland_2017, Wieland:2017zkf, DePaoli:2018erh, Oliveri:2019gvm}, but here we put the pieces together and squeeze all the physical content out of this formula. For Lorentz transformations, the corner potential yields symmetry charges satisfying an $\sll(2,\C)^S$ algebra\footnote{ $X^S$ means the set of maps $S\to X$.}, with the generators given by the corner 2-form
\be
E^{IJ}\stackrel{S}=\tE^Jn^I-\tE^{I}n^{J}+\beta(\te\wedge\te)^{IJ}.
\ee
We then find that there is an unambiguous description of the simplicity constraints on this corner phase space, and that the bulk and corner simplicity constraints are different in nature. More precisely, our parametrization of the bulk degrees of freedom with the GR symplectic structure removes the ambiguities associated to a choice of connection in the bulk, while an alternative and more symmetric parametrization of the corner symplectic potential (see Section \ref{sec:boundary with nIJ}) tells us how to introduce and make sense out of the corner simplicity constraints. In this way, the confusions which have subsisted so far completely wash away once the covariant phase space formalism is used to properly analyse the corner symplectic structure. For the first time, LQG without the time gauge is non-ambiguous. The same new parametrization naturally reveals the existence of an extra $\sll(2, \RR)_\para^S$ component of the corner symmetry algebra, which on-shell of the corner simplicity constraints is associated to the tangential components of the corner metric. This second main result of the paper, derived in Section \ref{sec:bmetric}, is
\be\la{res-2}
(ii)\q\{q_{ab}(x),q_{cd}(y)\}=-\f{1}{\beta}\big(q_{ac}\eps_{bd }+q_{bc}\eps_{ad }+q_{ad}\eps_{bc}+q_{bd}\eps_{ac}\big)(x)\delta^2(x,y),
\ee
where $\eps_{ab}$ is the totally skew Levi--Civita tensor. The corner area density is therefore related to the $\sll(2, \RR)_\para^S$ Casimir. Standard elements of representation theory together with the space-like nature of the corner metric yield straightforwardly a discrete spectrum for the corresponding area operator. The bracket \eqref{res-2} highlights the crucial role of the Barbero--Immirizi parameter in obtaining an area gap, consistently with the standard LQG description \cite{Rovelli:1994ge, Ashtekar:1996eg}. However, here we achieve this result while shifting the emphasis from the discrete holonomy-flux representation (and the use of the time gauge) to the continuum corner symmetry algebra \cite{Freidel:2015gpa, Freidel:2018pvm, Freidel:2019ees}. This result reconciles fundamental discreteness with Lorentz invariance, and it is a manifestation of an underlying more general geometrical structure which will be revealed and studied in the companion paper \cite{Edge-Mode-III}. It is interesting to point out that the discreteness of the area spectrum from the quantization of the corner phase space in the continuum has also been derived in \cite{Wieland:2017cmf}, although in a null context and using spinorial variables.

Finally, we show how the answer to question $(iii)$ is provided by the introduction of the concept of \textit{edge modes} as a set of fields $(\e^I,\JJ^{IJ}, \varphi)$ living at the corner and a priori independent of the (pull-back of the) bulk fields. These play a dual role: On the one hand, they are required to restore gauge invariance at the corner, while defining non-trivial corner symmetry charges. On the other hand, they are related to the pull-back of the bulk fields on the corner by a \textit{gauge frame} $\vphi\in\SL(2,\mathbb{C})^S$, which is an element of the corner symmetry group which dresses the corner flux \cite{Donnelly:2016auv}. As shown in Section \ref{sec:edge modes}, this gluing condition can be expressed in terms of the coframe field and the Lorentz generators as 
\be\la{res3}
(iii)\q\te_a^I\stackrel{S}{\simeq}\rho_a{}^b\e_b^J\varphi_J{}^I,\q\q E^{IJ}\stackrel{S}{\simeq}(\varphi^{-1}\JJ\varphi)^{IJ},
\ee
where an additional corner group element $\rho\in\SL(2,\R)^S$ is necessary for the coframe. Since the edge modes so introduced are non-commutative, their identification with the pull-back of the bulk flux and coframe fields encodes naturally the non-commutativity of these geometrical variables at the corner. Moreover, the introduction of the gauge frame $\varphi$ is here to insure that the gluing condition is a first class constraint invariant under the boundary symmetry algebra. The appearance of this gauge frame also opens up the possibility to understand how, even though we are not using a connection formulation to parametrize the bulk phase space in \eqref{res1} (as in standard LQG), the information about a gauge connection (through its holonomy group elements) can still be reconstructed from the corner data when considering the gluing of the coframe edge modes across subregions. This feature embodies the paradigm shift which we are proposing: The discrete bulk holonomy is no longer a constituent ingredient of the quantum geometry Hilbert space, but instead it emerges from the representation of the continuum ultra-local algebra of the corner symmetry charges. This interpretation of the gauge frame as a holonomy, which we just keep as an observation at this point, will be developed in \cite{Edge-Mode-III, Edge-Mode-IV}.

\begin{table}[h]
\centering
\begin{tabular}{|c|c|c|c|c|c|}
\hline
&\multicolumn{5}{|c|}{\textbf{Corner symmetries} $\mathfrak{g}_S$}\\
\hline
\textbf{Formulation of gravity}& $\mathrm{diff}(S)$ & $\sll(2,\mathbb{R})_\per$ & $\sll(2,\mathbb{R})_\para$ & $\;\mathfrak{su}(2)\;$ & $\mathrm{boosts}$ \\ 
\hline
Canonical general relativity (GR) & $\checkmark$ & & & & \\ 
\hline
Einstein--Hilbert (EH) & \checkmark &\checkmark & & & \\
\hline
Einstein--Cartan (EC) & \checkmark & & & & \checkmark \\
\hline
Einstein--Cartan--Holst (ECH) & \checkmark & &\checkmark & \checkmark & \checkmark \\
\hline
Einstein--Cartan--Holst + time gauge (ECHt) & \checkmark & &\checkmark & \checkmark & \\
\hline
\end{tabular}
\caption{Parts of the corner symmetry algebra which are non-trivially represented in various formulations of gravity. The $\sll(2,\RR)$ algebras denoted with $\perp$ and $\parallel$ are respectively associated with the metric components which are normal and tangent to the corner. The last two columns are the decomposition of the Lorentz algebra into boosts and rotations.}
\la{table}
\end{table}

The plan of the paper is as follows. In Section \ref{sec:BF-tetrad} we review the basic ingredients of BF theory and the tetrad formulation of gravity, including the pre-symplectic potential and the Hamiltonian charges. These are known results (see also \cite{Corichi:2013zza, Montesinos:2017epa, Frodden:2017qwh} for the study of gravitational charges in the first order formalism), on which we however give a new perspective in the rest of the paper.
 
In Section \ref{sec:3+1} we perform a detailed decomposition of all the geometrical quantities that play an important role in BF theory, including the simplicity constraints. This allows us to derive a bulk + corner decomposition of the BF potential in Section \ref{sec:BF potential}, and of the ECH potential in Section \ref{sec:ECH-potential}. This reveals the dynamical nature of the internal normal at the corner. In fact, the main theme of Section \ref{sec:ECH-potential} is to show how, by analogy with the metric case \cite{Freidel:2020xyx}, the ECH symplectic potential can be decomposed into a bulk term common to all other formulations of gravity, plus a corner term. The bulk piece yields the tangent diffeomorphism corner charge, with the spatial diffeomorphism constraint taking the form of a conservation law for the momentum aspect. While this understanding of the spatial diffeomorphism constraint is immediate in the metric case, in tetrad gravity this interpretation has been revealed and exploited only recently \cite{Freidel:2019ees, Freidel:2019ofr}. 

In Section \ref{sec:corner} we provide an alternative parametrization of the ECH corner potential in terms of a Lie algebra-valued horizontal 1-form. We use this parametrization to introduce the corner tangential metric and the corner simplicity constraints (which are the focus of the follow-up paper \cite{Edge-Mode-III}). The components of the corner metric satisfy an $\sll(2,\RR)^S$ algebra and represent additional corner Dirac observables. This establishes that the corner symmetry algebra of tetrad gravity is given by\footnote{We will see in \cite{Edge-Mode-III} that, in fact, there is an extra $\mathfrak{u}(1)$ component as the $\mathfrak{sl}(2,\mathbb{C})$ and $\sll(2,\RR)$ Casimirs are related by a balance equation. For the sake of the preliminary analysis of the simplicity constraints presented here, this aspect is not fundamental.} $\mathfrak{g}_S=\text{diff}(S)\ltimes\big(\mathfrak{sl}(2,\mathbb{C})^S\oplus\sll(2,\RR)_\para^S\big)$. We end the section by deriving a key result, which is the discreteness of the corner area spectrum from the continuum. 

After all this preparatory analysis we are ready to introduce the edge modes for ECH gravity in Section \ref{sec:edge modes}. This section is of a more conceptual nature, aiming at clarifying and reconciling within our general framework several contrasting statements found across the literature. This gives us the opportunity to set the stage for the next paper \cite{Edge-Mode-III}  in the series, where the edge modes of ECH gravity provide the conceptually cleanest setup to study the corner simplicity constraints in the continuum and classical theory. This analysis will reveal the advantage of the edge mode formalism in unraveling new geometrical structures of boundary degrees of freedom, which solves old puzzles in quantum geometry while opening new paths towards quantization. A concluding discussion is presented in Section \ref{conclusions}.

We have included a long list of appendices containing details of various calculations. Our notations and conventions are gathered in Appendix \ref{appendix1}. Appendices \ref{gravityeoms}, \ref{appendix3} and \ref{AppD} collect several proofs and details of calculations used in the main text. An alternative decomposition of the ECH potential is presented in Appendix \ref{appendix:AE potential}. Further details on the Hamiltonian diffeomorphism charges are included in Appendix \ref{appendix:relative diffeo}.  Proof of the first class nature of the gluing condition \eqref{res3} is given in Appendix \ref{B gluing}.

\section{BF theory and tetrad formulation of gravity}
\la{sec:BF-tetrad}

In this section, following \cite{Freidel:2020xyx}, we introduce notations and review basic facts about BF theory and the Einstein--Cartan--Holst formulation of gravity. The reason for doing this is that we are going to derive in \eqref{BF decomposition} a new decomposition of the BF symplectic potential (into bulk and corner pieces), from which the decomposition \eqref{ECH decomposition} of the ECH potential will immediately follow upon imposing the simplicity constraints (see \cite{Celada:2016jdt} for a review of BF gravity).

\subsection{BF theory}

In terms of a Lorentz tensor 2-form $B^{IJ}$ and a Lorentz connection 1-form $\omega^{IJ}$ with curvature $F^{IJ}=\rd\omega^{IJ}+\omega^I{}_K\wedge\omega^{KJ}$, BF theory is defined by the Lagrangian
\be
L_\BF=\f{1}{2}B_{IJ}\wedge F^{IJ}.
\ee
The equations of motion are the flatness and Gauss equations
\be
F^{IJ}\approx0,
\q
T^{IJ}\coloneqq\rd_\omega B^{IJ}\approx0.
\ee
The Bianchi identities
\be
\rd_\omega F^{IJ}=0,
\q
\rd_\omega T^{IJ}=[F,T]^{IJ}, 
\ee
signal the presence of two sets of gauge invariances,\footnote{The interplay between Bianchi identities and gauge symmetries goes back to Noether \cite{Noether}, and results in the conservation laws
\be
\rd(\alpha^{IJ}T_{IJ})=\delta_\alpha\omega_{IJ}T^{IJ}+\delta_\alpha B^{IJ}F_{IJ},
\q
\rd(\phi^{IJ}F_{IJ})=\delta_\phi\omega_{IJ}T^{IJ}+\delta_\phi B^{IJ}F_{IJ}.
\ee}
labelled by a Lie algebra-valued scalar $\alpha^{IJ}$ and a Lie algebra-valued 1-form $\phi^{IJ}$, and acting as
\be
\delta_\alpha B^{IJ}=[B,\alpha]^{IJ},
\q
\delta_\alpha\omega^{IJ}=\rd_\omega\alpha^{IJ},
\q
\delta_\phi B^{IJ}=\rd_\omega\phi^{IJ},
\q
\delta_\phi\omega^{IJ}=0.
\ee
Finally, the symplectic potential of BF theory, associated with a codimension-1 manifold $\Sigma$, is simply given by
\be\la{bare BF potential}
\Theta_\BF=\f{1}{2}\int_\Sigma B_{IJ}\wedge\delta\omega^{IJ}.
\ee
We  perform its bulk + corner decomposition in Section \ref{sec:BF potential}.

\subsection{Einstein--Cartan--Holst gravity}
\la{subsec:tetrad}

In the first order tetrad formulation of gravity, the basic fields are an $\mathbb{R}^4$-valued form, or coframe field $e^I=\rd x^\mu e_\mu^I$, with inverse $\hat{e}_I=e_I^\mu\partial_\mu$, and a Lorentz connection $\omega^{IJ}$. In terms of the coframe field, the spacetime metric is $g_{\mu\nu}=e_\mu^Ie_\nu^J\eta_{IJ}$, where $\eta_{IJ}=\text{diag}(-1,1,1,1)$. Coframes and their dual frames are related by $e_I^\mu=g^{\mu\nu}\eta_{IJ}e_\nu^J$.

The Einstein--Cartan--Holst (ECH) Lagrangian is
\be\la{ECH Lagrangian}
L_\ECH=\f{1}{2}E_{IJ}\wedge F^{IJ},
\q
E_{IJ}[e]\coloneqq(*+\beta)(e\wedge e)_{IJ}.
\ee
The duality map acting on the Lie algebra is defined as $(*M)_{IJ}=\tfrac12{\eps_{IJ}}^{KL}M_{KL}$, and we use the notation $(e\wedge e)^{IJ}\coloneqq e^I\wedge e^J$. We refer the reader to Appendix \ref{appendix1} for the rest of our notations and conventions, as well as some useful formulas. The parameter $\gamma=\beta^{-1} $ is the so-called Barbero--Immirzi parameter. It corresponds to a shift of the Lagrangian by the topological Holst term\footnote{There are other topological terms which one can add to the Lagrangian, corresponding to the Pontrjagin, Euler, and Nieh--Yan classes \cite{Nieh-Yan,Freidel:2005ak,Rezende_2009,Corichi:2013zza,Corichi:2016zac}. We will come back to these terms in their influence on the potential in future work.} \cite{Holst:1995pc}. The ECH Lagrangian is obtained from the BF one after imposition of the {\it simplicity constraints}
\be
(*B)^{IJ}-\beta B^{IJ}=-(1+\beta^2)(e\wedge e)^{IJ}\q\Rightarrow\q B^{IJ}=E^{IJ}[e].
\ee
These constraints will be analyzed and decomposed in Section \ref{sec:simplicity} below. Their corner counterpart and their quantization will be the focus of the companion papers \cite{Edge-Mode-III, Edge-Mode-IV}, but we present a preliminary analysis of the corner simplicity constraints already here in Section \ref{sec:boundary simplicity}.

The equations of motion obtained by varying the ECH Lagrangian with respect to $e^I$ and $\omega^{IJ}$ respectively are given by
\be\la{EC EOMs}
G^I\coloneqq(*+\beta)F^{IJ}\wedge e_J\approx0,\q T_{IJ}\coloneqq(*+\beta)\big(\rd_\omega e_{[I}\wedge e_{J]}\big)\approx0,
\ee
with $G^I$ the Einstein tensor in tetrad variables. When the coframe is invertible, the second equation is equivalent to the vanishing of the torsion $T^I\coloneqq\rd_\omega e^I$. These first order equations of motion satisfy two Bianchi identities, namely
\be 
\rd_\omega T_{IJ}=e_{[I}\wedge G_{J]},
\q
\xi_I\rd_\omega G^I=\xi\ip T_I\wedge G^I+\xi\ip F^{IJ}\wedge T_{IJ},
\ee
where $\xi_I=\xi\ip e_I$. These identities signal the presence of two sets of gauge invariances. They correspond to internal Lorentz transformations, which are labelled by a Lie algebra valued scalar $\alpha^{IJ}$, and diffeomorphisms, which are labelled by a vector field $\xi$. Their action on the fields is
\be\la{gauge transfos}
\delta_\alpha e^I=-\alpha^I{}_Je^J,
\q
\delta_\alpha\omega=\rd_\omega\alpha^{IJ},
\q
\delta_\xi e^I=\CL_\xi e^I,
\q
\delta_\xi\omega^{IJ}=\CL_\xi\omega^{IJ},
\ee
where $\CL_\xi(\cdot)=\rd(\xi\ip\cdot)+\xi\ip(\rd\,\cdot)$ is the Lie derivative. The charges associated with these transformations have been studied in \cite{Freidel:2020xyx}. Here we have another look at them once the symplectic potential ECH has been decomposed into bulk and corner components.

The symplectic potential for tetrad gravity which we are going to study in the rest of this paper, and which follows from the Einstein--Cartan--Holst Lagrangian, is
\be\la{bare ECH potential} 
\Theta_\ECH=\f{1}{2}\int_\Sigma E_{IJ}\wedge\delta\omega^{IJ}.
\ee
For completeness we recall that the corner Hamiltonian charges associated with Lorentz transformations and diffeomorphisms are
\be\la{ECH charges}
\CH^S_\ECH[\alpha]=\f{1}{2}\int_S\alpha_{IJ}E^{IJ},
\q
\CH^S_\ECH[\xi]=\f{1}{2}\int_S\xi\ip\gamma_{IJ}E^{IJ},
\ee
where $\gamma^{IJ}[e]$ is the torsionless Lorentz connection compatible with $e^I$. On-shell of the Gauss and diffeomorphism constraints, these charges satisfy a non-commutative algebra, which is the corner algebra $\mathrm{diff}(S)\ltimes \sll(2,C)^S$ under study \cite{Edge-Mode-I}
\begin{subequations}\label{non-c}
\be
\{\CH^S_\ECH[\alpha],\CH^S_\ECH[\beta]\}&\simeq\CH^S_\ECH\big[[\alpha,\beta]\big],\\
\{\CH^S_\ECH[\xi],\CH^S_\ECH[\alpha]\}&\simeq\CH^S_\ECH[{\cal L}_\xi\alpha],\\
\{\CH^S_\ECH[\xi],\CH^S_\ECH[\xi']\}&\simeq\CH^S_\ECH\big[[\xi,\xi']_{\mathrm{Lie}}\big].
\ee
\end{subequations}

One can clearly obtain the potential $\Theta_\ECH$ from $\Theta_\BF$ by imposing the simplicity constraints. Our task is now to explain how this can be achieved when the potentials are decomposed into bulk and corner components. This  also shows the explicit relationship between $\Theta_\ECH$ and the universal bulk piece $\Theta_\GR$. In order to get these results, we now need to understand how all the various quantities which have appeared so far can be decomposed geometrically on the slice $\Sigma$.

\subsection{A new look at canonical analysis}

The role of the simplicity constraints in reducing topological BF theory to Einstein--Cartan--Holst gravity has been extensively studied in the literature, as well as the canonical structure of the ECH Lagrangian \eqref{ECH Lagrangian} (see references in the next paragraph). However, this was done almost exclusively using Dirac's algorithm of Hamiltonian analysis. There exists a second canonical way of studying the phase space of a classical theory, which is to use the covariant phase space formalism as we do here. These two possibilities differ in the sense that the former uses a separation of the constraints into first/second class, as well as primary/secondary/etc\dots, while the latter is simply an on-shell formalism which is agnostic about such a separation. It is therefore important to explain at this point where our treatment stands with respect to the already existing literature, and how it enables us to solve some long standing puzzles in canonical LQG and spin foam models.

The Hamiltonian analysis of Einstein--Cartan gravity was performed in \cite{Ashtekar:1991hf,Romano_1993,Peld_n_1994}. It leads to a phase space parametrized by the ADM canonical pair (given by the induced metric on the slice and its conjugate momentum associated to the extrinsic curvature of the slice), or equivalently its first order tetrad analog. The classical starting point of LQG, however, is a parametrization of the phase space in terms of an $\su(2)$ connection and the conjugated densitized triad, traditionally called the \textit{flux}. This was initially derived as a canonical transformation from the ADM phase space \cite{Barbero:1994an,Barbero:1994ap}, and then from the Hamiltonian analysis of the Einstein--Cartan--Holst   Lagrangian \cite{Holst:1995pc}. These derivations however use the time gauge, which amounts to fixing the internal normal $n^I$, and therefore reduces the internal gauge group from $\SL(2,\C)$ to the $\SU(2)$ stabilizing $n^I$. It was suggested that this gauge choice was at the origin of the discreteness of the area spectrum in LQG \cite{Alexandrov:2001pa,Alexandrov:2010un}, and the source of difficulties when trying to match canonical LQG with the covariant spin foam approach \cite{Alexandrov:2007pq,Alexandrov:2008da,Alexandrov:2010pg,Alexandrov:2010un,Alexandrov:2011ab}. This has motivated the study of the canonical theory without the time gauge. When analyzing the ECH Lagrangian without the time gauge, complications arise due to the presence of second class constraints,\footnote{Second class constraints are also present in the time gauge, but in this case they can be handled easily and lead to an ambiguous parametrization of the phase space \cite{Holst:1995pc,Geiller_2013}.} which are precisely the primary simplicity constraints and their conjugated secondary constraints \cite{BarroseSa:2000vx,Alexandrov:2006wt,Alexandrov:2011ab}. These can be dealt with either by using the Dirac bracket \cite{Alexandrov:2000jw,Alexandrov:2002br},  by working with an explicit solution of the constraints \cite{Cianfrani:2008zv,Geiller:2011bh,Montesinos_2019}, by working with an explicit solution of the constraints \cite{Cianfrani:2008zv,Geiller:2011bh,Montesinos_2019}
or by adding variables to promote the constraints to a first class set \cite{Montesinos:2019mxg,Montesinos:2019bgd}. In all cases, one arrives at the conclusion that the choice of Lorentz connection configuration variable is ambiguous in the bulk \cite{Alexandrov:2001wt,Alexandrov:2011ab}, an observation which has fueled the  discussion on the ambiguities of the imposition of the spin foam simplicity constraints.\footnote{In the Hamiltonian analysis the simplicity constraints come in pairs given by primary and secondary constraints, and these latter depend on which connection variable is chosen in order to parametrize the phase space.} Obviously this also translates into ambiguities in the quantum theory.

Therefore, some of the main questions which remain open from the point of view of the Hamiltonian analysis of the simplicity constraints in the bulk are: Does the discreteness of the area spectrum survive the relaxation of the time gauge and is it compatible with Lorentz invariance? Are the simplicity constraints properly imposed in spin foam models? Indications that the answer to the first question is positive are given in \cite{Geiller:2011bh,Wieland:2017cmf} (see also \cite{Dittrich:2007th,Rovelli:2010ed}). The second question has been the source of much debate \cite{Alexandrov:2008da,Anza:2014tea,Dittrich:2008ar,Dittrich:2010ey,Dupuis:2010iq,Geiller:2011aa,Livine:2007ya}. In our view, confusion in this debate is deeply rooted in a more fundamental puzzle of the LQG framework, which consists of the incompatibility between  the bulk and the corner phase space   canonical structures, as pointed out in the introductory Section \ref{sec:questions}.


The rest of this paper is devoted to addressing these issues. Our new treatment is based, at the conceptual level, on the shift of focus from the bulk connection to the corner coframe field, and, at the technical level, on the analysis of tetrad gravity by means of the covariant phase space formalism. The latter 
 being an on-shell construction, all that is required in order to deal with the second class constraints (which after all are just the canonical decomposition of the torsion equations) is to properly decompose the connection and to impose the equations of motion. This is what we do just below.
Let us now dive step by step into the technicalities of this construction. For this we perform a decomposition of the various geometrical quantities using the spacetime and internal normals.

\section{$\boldsymbol{3+1}$ decompositions}
\la{sec:3+1}

In this section we introduce all the geometrical tools necessary in order to decompose the symplectic potentials \eqref{bare BF potential} and \eqref{bare ECH potential} respectively associated with topological BF theory and ECH gravity. This requires to decompose the constraints, equations of motions, Lorentz tensors, and connections. These decompositions reveal the precise form of the bulk and corner components of the potentials, and their geometrical interpretation. They also clarify the role of the bulk simplicity constraints.

We start by recalling some standard material. As usual, the $3+1$ decomposition relies on a foliation of the spacetime by codimension-1 submanifolds $\Sigma$. This foliation defines a normal 1-form $\un=n_\mu\rd x^\mu$ and the dual normal vector $\hat{n}=n^\mu\partial_\mu$. Given a coframe $e^I=\rd x^\mu e_\mu^I $ and dual frame $\hat{e}_I=e_I^\mu\partial_\mu$, we can introduce an internal normal $n^I=\hat{n}\ip e^I$ such that
\be
\un=e^In_I,\q\hat{n}=\hat{e}_In^I,\q n^In_I=n^\mu n_\mu=\sigma,
\ee
with $\sigma=-1$ for a time-like normal and $\sigma=+1$ for a space-like one (whenever such a choice is made in the following, it will be explicitly said).

With the help of these normals we can then introduce a new coframe field
\be\la{tilde e}
\te^I_\mu\coloneqq e^I_\mu-\sigma n_\mu n^I.
\ee
This form is both \textit{tangential} in the sense $\te^In_I=0$, and \textit{horizontal} in the sense $\hat{n}\ip\te^I=n^\mu\te_\mu^I=0$. It furthermore defines the induced metric on $\Sigma$ as
\be
\tilde{g}_{\mu\nu}\coloneqq\te_\mu^I\te_\nu^J\eta_{IJ}=g_{\mu\nu}-\sigma n_\mu n_\nu.
\ee
Note that it is important to differentiate between the normals $\un,\hat{n},n^I$ because they have a different behavior under field variations. One usually assumes that the normal form $\un$ is kinematical, i.e. independent of the metric except for its normalization. This means that we impose that $\delta\un\propto\un$ under a field variation. This condition implies that the field variation preserves the chosen foliation. It is important, however, to appreciate that the normal vectors $\hat{n}$ and $n^I$ are both phase space variables which possess non-trivial field variations. Indeed, we have
\be
\te^I\delta n_I=-\delta\te^In_I,
\q
\te_\mu^I\delta n^\mu=-\delta\te_\mu^In^\mu,
\ee
which follows directly from the fact that $\te$ is both tangential and horizontal. The inclusion of the internal normal $n^I$ in the phase space is key to the construction which we present here.

In BF theory the fundamental field is the Lie algebra-valued 2-form $B^{IJ}$, and there is no intrinsic notion of coframe field. However, it turns out that once we chose an internal normal vector $n^I$ we can define a tangential boost coframe $\tb^I$ and a tangential spin coframe $\ts^I$. Decomposing $B^{IJ}$ in terms of these coframes enables us to decompose the BF potential in a form very similar to the ECH potential. Then, the simplicity constraints have an elegant rewriting as a relationship between the coframes $(\tb^I,\ts^I)$ and the gravitational coframe $\te^I$.

Two important notions in what follows are the decompositions of Lorentz and spacetime indices into normal/tangential and vertical/horizontal components respectively. The first decomposition refers to the internal Lorentz indices, while the second one to spacetime differential form indices. Let us now introduce these decompositions

\subsection{Normal/tangential decomposition}

Let us start by discussing the decomposition into normal and tangential components. We first use it for tensors like $B^{IJ}$ and $E^{IJ}$, then for connections like $\omega^{IJ}$, and then for the Gauss constraint, where it corresponds to a decomposition into boosts and rotations.

\subsubsection{Decomposition of Lorentz tensors}

Given the internal normal $n^I$ and a Lorentz tensor $M^{IJ}$, we define its normal and tangential components as
\be\la{normal/tangential parts}
M_\per^I\coloneqq M^{IJ}n_J,
\q
M_\para^I\coloneqq(*M)^{IJ}n_J.
\ee
With this we can then decompose the Lorentz tensor as
\be\la{tensor decomposition}
M^{IJ}=2\sigma M_\per^{[I}n^{J]}-\sigma\teps^{IJ}{}_{K}M_\para^K,
\ee
where we have introduced the induced epsilon tensor $\teps_{IJK}\coloneqq\eps_{IJKL}n^L$, and we recall that antisymmetrization of indices is defined with a factor $1/2$. Under duality $M\rightarrow*M$, we have that $M_\per\rightarrow M_\para$ and $M_\para\rightarrow-M_\per$.

We make extensive use of the decomposition of the 2-form $B^{IJ}$. Since it appears as the generator (or charge) of Lorentz transformation, we can naturally understand its tangential and normal components as generators of rotation and boost. In order to reflect this we therefore adopt the notation
\be
B^I\coloneqq B^{IJ}n_J,
\q
S^I\coloneqq(*B)^{IJ}n_J,
\ee
where $B^I$ stands for boost and $S^I$ for spin. In terms of these components, we will repeatedly use the decomposition
\be\la{tensor decomposition B}
\boxed{\quad B^{IJ}=2\sigma B^{[I}n^{J]}-\sigma\teps^{IJ}{}_{K}S^K.\quad}
\ee
This is  particularly  important for the decomposition of the BF potential and the rewriting of the simplicity constraints.

\subsubsection{Decomposition of connections}

Both BF theory and ECH gravity feature the Lorentz connection $\omega^{IJ}$, whose decomposition is a central ingredient of what follows. The spacetime Lorentz connection can be decomposed as
\be\la{decomposition of omega}
\boxed{\quad\omega^{IJ}=\Gamma^{IJ}+2\sigma K^{[I}n^{J]}.\quad}
\ee
In this decomposition, the only requirement which we impose is
\be\la{covariant Gamma of n}
\rd_\Gamma n^I=0.
\ee
This in turn implies that
\be
K^I=\rd_\omega n^I.
\ee
This shows that $K^I$ is a tangential Lorentz vector since $K^In_I=0$. As the space of Lorentz connections is an affine space, the fact that $K^I$ is a tensor implies that $\Gamma^{IJ}$ is a Lorentz connection. More precisely, it is the connection which preserves $n^I$. Then, the fact that $\rd_\Gamma n^I=0$ means that its curvature tensor $R[\Gamma]=\rd\Gamma+\Gamma\wedge\Gamma$ is purely tangential, i.e. such that $R^{IJ}n_J=0$.

One should be careful when comparing this decomposition of the connection with the decomposition of tensors as in \eqref{tensor decomposition}. Indeed, one can see that $K^I$ is \textit{not} equal to the normal part of the connection, since $\omega^I_\per=\omega^{IJ}n_J=K^I-\rd n^I$. However, we have that the tangential part of $\Gamma^{IJ}$ \textit{is} the tangential part of the Lorentz connection, i.e. $\omega_\para^I=(*\omega)^{IJ}n_J=\Gamma_\para^I$. In Appendix \ref{appendix:gauge} we give for completeness the decomposition of the Lorentz gauge transformations acting on $K^I$ and $\Gamma^{IJ}$. This shows as expected that the former transforms as a tensor and the latter as a Lorentz connection.

Finally, the decomposition of the connection implies that its curvature tensor decomposes as
\be\la{components of Lorentz F}
F^{IJ}=R^{IJ}(\Gamma)-\sigma(K\wedge K)^{IJ}+2\sigma\rd_\Gamma K^{[I}n^{J]}.
\ee
We can read from this the normal and tangential components, which are respectively $F_\per^I=\rd_\Gamma K^I$ and $F_\para^{IJ}=R^{IJ}(\Gamma)-\sigma(K\wedge K)^{IJ}$. We use this later on when decomposing the Einstein equations.

\subsubsection{Boost/rotation decomposition of the Gauss constraint}

It is now useful to apply the tangential/normal decomposition to the Gauss constraint $T^{IJ}=\rd_\omega B^{IJ}$, which leads to a rotational and a boost Gauss laws. First, using \eqref{tensor decomposition B} and \eqref{decomposition of omega} leads to
\be
\rd_\omega B^{IJ}=\rd_\Gamma B^{IJ}+2\sigma B^{[I}\wedge K^{J]}+2(K\times S)^{[I}n^{J]},
\ee
where we have used the cross product $(M\times N)^I\coloneqq{\teps^I}_{JK}M^J\wedge N^K$. Then, using \eqref{tensor decomposition B} once again, and the fact that $\rd_\Gamma\teps_{IJK}=0$, we can further decompose
\be
\rd_\Gamma B^{IJ}=2\sigma\rd_\Gamma B^{[I}n^{J]}- \sigma \teps^{IJ}{}_K \rd_\Gamma S^K .
\ee
This implies that the boost and rotation components of the Gauss constraint, given respectively by $\CB^I\coloneqq T^{IJ}n_J$ and $\CR^I\coloneqq(*T)^{IJ}n_J$, are equal to
\be\la{rotation boost}
\CB^I=\rd_\Gamma B^I+\sigma(K\times S)^I,
\q
\CR^I=\rd_\Gamma S^I-\sigma(K\times B)^I.
\ee
The boost component  makes an appearance  in the bulk piece of the BF symplectic potential (when written off-shell). Furthermore, using the simplicity constraints in \eqref{rotation boost} will give a boost/rotation decomposition of the torsion equations of motion of ECH gravity.

\subsection{Horizontal/vertical decomposition}

The presence of the normal form $\un$ allows us to decompose any spacetime form $\alpha$ into horizontal and vertical components. The vertical component of the form $\alpha$ is the form $\alpha_n\coloneqq\hat{n}\ip\alpha$ obtained by contraction with the normal vector. Therefore, a form $\alpha$ is said to be horizontal when its vertical component vanishes, i.e. when $\hat{n}\ip\alpha=0$. The horizontal component $\tilde{\alpha}$ is the component which survives the pull-back on $\Sigma$, and is denoted by $\alpha\stackrel{\Sigma}{=}\tilde{\alpha}$. With this, any form $\alpha$ can be decomposed into horizontal and vertical components as
\be\la{horizontal/vertical}
\alpha=\tilde{\alpha}+\sigma\un\wedge\alpha_n,
\q
\alpha_n\coloneqq\hat{n}\ip\alpha.
\ee
In what follows, forms with a tilde will always be horizontal. Notice that the coframe field $e^I$ is special in the sense that its horizontal component is also tangential, as explained below \eqref{tilde e}, although this is not true for a general form.

We can also decompose ordinary and covariant differentials of forms. For example, for an horizontal form $\tilde{\alpha}$ the covariant derivative satisfies
\be
\rd_\Gamma\tilde{\alpha}=\tilde{\rd}_\Gamma\tilde{\alpha}+\sigma\un\wedge\CL^{\va\Gamma}_{\hat{n}}\tilde{\alpha},
\ee
where $\tilde{\rd}$ is the pull-back differential on $\Sigma$, and where we have introduced the covariant Lie derivative $\CL^{\va\Gamma}_\xi\tilde{\alpha}^I\coloneqq\xi\ip(\rd_\Gamma\tilde{\alpha}^I)+\rd_\Gamma(\xi\ip\tilde{\alpha}^I)$. This covariant Lie derivative is such that $\CL^{\va\Gamma}_\xi n^I=0$. We also have
\be
\rd\un=\tilde{\rd}\un+\sigma\un\wedge\CL_{\hat{n}}\un.
\ee
This decomposition can be applied to the tensor appearing in the decomposition of the connection. In this case we have
\be\la{decomposition of K}
K^I=\tK^I+\sigma\un\wedge K_n^I,
\ee
where $\tK^I $ is the \textit{extrinsic curvature} form and $K_n^I$ can be identified with the \textit{acceleration}. Using the fact that $\CL_\xi\omega=\xi\ip F+\rd_\omega(\xi\ip\omega)$, the vertical component of the curvature can be expressed in terms of the Lie derivative of the connection along the vector $\hat{n}$ as
\be
F_n^{IJ}=\CL_{\hat{n}}\omega^{IJ}-\rd_\omega\omega_n^{IJ}.
\ee

\subsection{BF coframes}

Let us now focus again on the Lie algebra-valued 2-form $B^{IJ}$. We can decompose it in both tangential/normal and horizontal/vertical components, but then also combine these decompositions. Then, both the normal (boost) and the tangential (spin) parts can be decomposed into horizontal and vertical components as
\be\la{B S decompositions}
B^I=\tilde{B}^I+\sigma\un\wedge B_n^I,
\q
S^I=\tilde{S}^I+\sigma\un\wedge S_n^I.
\ee
With these two decompositions, we have rewritten the 36 components of the 2-form $B^{IJ}$ in terms of the $18+18$ components $(B^I,S^I)$, and then each of these (say, for $B^I$) 18 components in terms of the $9+9$ horizontal/vertical components $(\tilde{B}^I,B_n^I)$. The object $B_n^I$ is a vector-valued vertical 1-form, and can be thought of as a the vertical component of a coframe, which we call the \textit{normal boost coframe}. On the other hand
$\tB^I$ is a vector-valued horizontal 2-form. It is important to appreciate that it is possible to generically\footnote{Provided that $\tB^I$ satisfies the non-degeneracy condition $\eps^{abc}\teps_{IJK}\tB_{ab}^I\tB_{cd}^J\neq0$.} decompose $\tB^I$ as a cross product of horizontal 1-forms $\tb^I$ by writing \be\la{Bb}
\tB^I=\f{1}{2}(\tb\times\tb)^I,
\q
(\tb\wedge\tb)^{IJ}=-\sigma{\teps^{IJ}}_K\tB^K.
\ee
Indeed, here we are simply trading the 9 components of $\tB^I$ for the 9 components of $\tb^I$. We call $\tb^I$ the \textit{tangential boost coframe}. Similarly we can rewrite $\tilde{S}^I$ as
\be\la{Ss}
\tilde{S}^I=\f{1}{2}(\ts\times\ts)^I,
\q
(\ts\wedge\ts)^{IJ}=-\sigma{\teps^{IJ}}_K\tS^K,
\ee
in terms of a \textit{tangential spin coframe} $\ts^I$. Notice that these decompositions hold \textit{before} imposing the simplicity constraints relating $B^{IJ}$ to the gravitational $E^{IJ}[e]$. Here we are still in BF theory, which is why there are two coframes\footnote{This can also be understood in terms of the Urbantke metrics of BF theory \cite{Freidel_2012}.}. We show below that the simplicity constraints have an elegant interpretation as relating $\tb^I$ and $\ts^I$ with the gravitational coframe $\te^I$.

This completes our tangential/normal and horizontal/vertical decomposition of the $B$ field in terms of two boost coframes $(\tb,B_n)$ and two spin coframes $(\ts,S_n)$. We use this decomposition in Section \ref{sec:BF potential} to rewrite the potential of BF theory.

\subsection{Equations of motion} 

We now present the decomposition of the equations of motion of Einstein--Cartan--Holst gravity. These are the torsion equation\footnote{The torsion equation can be imposed alone, which we denote by $\simeq$, while the Einstein equation requires to also impose the torsion, and is therefore denoted by $\approx$.} $T^I=\rd_\omega e^I\simeq0$ and the Einstein equation $G^I\approx0$ introduced in \eqref{EC EOMs}. The details are given in Appendix \ref{gravityeoms}.

The torsion equation can be decomposed into normal $T_\perp=T^In_I$ and tangential $T^I_\para$ parts, and equation \eqref{torsion decomposition} gives its decomposition into horizontal and vertical components. The horizontal components give the \textit{constraints}, while the vertical components are the \textit{evolutions equations}. More precisely, the tangential and normal horizontal components are respectively given by
\be\la{Tconst}
\tilde{\rd}_\Gamma\te^I\simeq0,
\q
\te_I\wedge\tK^I\simeq 0.
\ee
The first equation establishes that $\tilde{\Gamma}$ is the spin connection associated with $\te$. The second equation can be understood as the condition 
that the extrinsic curvature tensor is symmetric, i.e. $\tK^{[IJ]}=0$ where $\tK^I=\tK^{IJ}\te_J$. The vertical components $\hat{n}\ip T_\para^I$ and $\hat{n}\ip T_\perp$ define the evolution equations
\be 
\CL^{\va\Gamma}_{\hat{n}}\te^I\simeq\tK^I,
\q
\te_I  K_n^I\simeq\CL_{\hat{n}}\un,
\ee
where $\CL^{\va\Gamma}_\xi$ is the covariant Lie derivative associated with $\Gamma$. The first relation tells us that the pull-back of $K^I$ can be understood as the extrinsic curvature, i.e. the normal derivative of the induced coframe.

Using the normal and tangential components \eqref{normal/tangent F} of the Lorentz curvature tensor $F^{IJ}$, we can similarly decompose the Einstein tensor $G^I=(*+\beta)F^{IJ}\wedge e_J$ into normal $G_\perp=G^In_I$ and tangential $G^I_\para$ components. We then denote by $C$ the horizontal component of $G_\perp$, and by $C^I$ the horizontal component of $G^I_\para$. The quantities $C$ and $C^I$ are the constraints when $\sigma=-1$, 
and they are boundary evolution equations when $\sigma=+1$. Explicitly, they are given by
\be\la{conservation P}
C^I\simeq\tilde{\rd}_\Gamma\tP^I,
\q
C=-\tilde{R}^I(\tilde{\Gamma})\wedge\te_I+\f{\sigma}{2}(\tK\times\tK)^I\wedge\te_I-\beta\tilde{\rd}_\Gamma\tK^I\wedge\te_I,
\ee
where we had to use the torsion equations to rewrite $C^I$. These two expressions are nothing but the spatial diffeomorphism constraint and the Hamiltonian constraint. The spatial diffeomorphism constraint is here nicely expressed as a conservation equation for the momentum aspect \cite{Freidel:2019ees}
\be\la{momentum aspect}
\tP^I\coloneqq-\sigma(\tK\times\te)^I.
\ee
Note that the last term of $C$ vanishes when the torsion constraints \eqref{Tconst} are satisfied. The vertical components $\hat{n}\ip G_\perp$ and $\hat{n}\ip G_\para$ are the evolutions equations.

\subsection{Bulk simplicity constraints}
\la{sec:simplicity}

In their most elementary form, the simplicity constraints taking us from BF theory to ECH gravity are just the requirement that
\be\la{initial simplicity}
B^{IJ}=E^{IJ}[e],
\q
E^{IJ}[e]=(*+\beta)(e\wedge e)^{IJ}.
\ee
It is now natural to decompose both sides of this relation in terms of horizontal/vertical and normal/tangential components, and thereby express the simplicity constraints as relations between the various components of $B^{IJ}$ and $E^{IJ}$.

Let us start by decomposing the gravitational 2-form $E^{IJ}$. Applying the tangential/normal decomposition to the wedge product of the coframe fields, we get
\be
(e\wedge e)^I_\per=(e\wedge e)^{IJ}n_J=\te^I\wedge\un,
\q
(e\wedge e)^I_\para=*(e\wedge e)^{IJ}n_J=\f{1}{2}(\te\times\te)^I.
\ee
We now introduce the horizontal 2-form $\tE^I$ (called the flux in LQG) defined as
\be\la{definition E}
\tE^I\coloneqq\f{1}{2}(\te\times\te)^I,
\q
(\te\wedge\te)^{IJ}=-\sigma{\teps^{IJ}}_K\tE^K,
\q
*(\te\wedge\te)^{IJ}=2\sigma\tE^{[I}n^{J]}.
\ee
From this definition, one can see that $\tE^I$ is also a tangential form, i.e. $\tE^In_I=0$. Using these ingredients in \eqref{tensor decomposition} tells us that
\be
(e\wedge e)^{IJ}=2\sigma\te^{[I}n^{J]}\wedge\un-\sigma{\teps^{IJ}}_K\tE^K\,,
\q
*(e\wedge e)^{IJ}=2\sigma\tE^{[I}n^{J]}+\sigma{\teps^{IJ}}_K\te^K\wedge\un\,,
\ee
and we find that the decomposition of the gravitational 2-form is
\be
E^{IJ}=\sigma\big(2\tE^{[I}n^{J]}-\beta{\teps^{IJ}}_K\tE^K\big)-\sigma\un\wedge\big(2\beta\te^{[I}n^{J]}+{\teps^{IJ}}_K\te^K\big).
\ee
This means in particular that the pull-back to $\Sigma$ of $E^{IJ}$ is given by
\be\la{E on Sigma}
\boxed{\quad\tE^{IJ}=2\sigma\tE^{[I}n^{J]}+\beta(\te\wedge\te)^{IJ}=\sigma\big(2\tE^{[I}n^{J]}-\beta{\teps^{IJ}}_K\tE^K\big)\quad}
\ee
and it satisfies $\tE^{IJ}n_J=\tE^I$.

Similarly, using the decompositions \eqref{tensor decomposition B} and \eqref{B S decompositions} of $B^{IJ}$ in terms of boost and spin coframes yields
\be
B^{IJ}=\sigma\big(2\tB^{[I}n^{J]}-{\teps^{IJ}}_K\tS^K\big)+\un\wedge\big(2B^{[I}_nn^{J]}-{\teps^{IJ}}_KS^K_n\big).
\ee
The simplicity constraints $B^{IJ}=E^{IJ}$ can therefore be written in terms of the boost and spin components as
\be\la{simplicity 1}
\tB^I=\tE^I,
\q
\tS^I=\beta\tE^I,
\q
B_n^I=-\sigma\beta\te^I,
\q
S_n^I=\sigma\te^I.
\ee
One sees that the simplicity constraints identify the tangential boost form with the gravitational flux. To go from the BF symplectic potential \eqref{BF decomposition} to the ECH symplectic potential \eqref{ECH decomposition}, it will be sufficient to focus only on the fact that the simplicity constraints identify
\be\la{simplicity 2}
\tB^I=\tE^I,
\q
\ts^I=\sqrt{\beta}\te^I.
\ee
Notice that the actual names of the coframes are irrelevant: What matters is that BF theory has \textit{two} coframes while ECH gravity only has \textit{one}. The bulk simplicity constraints say that the two coframes of BF theory are in fact not independent, but proportional to each other through $\beta$ (if $\beta=0$, the simplicity constraints amount to killing one of the BF coframes). For this reason, we could have chosen from the onset a ``notational gauge'' in which the pair of BF coframes is either $(\tb^I\equiv\te^I,\ts^I)$ or $(\tb^I,\ts^I\equiv\displaystyle\sqrt{\beta}\te^I)$. We have chosen not to do so at this point in order not to mix notations between BF and ECH and introduce possible sources of confusion. We however choose the latter notational shortcut in Sections \ref{sec:corner} and \ref{sec:edge modes}.

An important word of caution must be mentioned at this point, and has to do with the difference between the bulk and corner simplicity constraints. In the bulk, reducing topological BF theory to ECH gravity requires to impose $B^{IJ}=E^{IJ}$, which as we have shown translates into the various identifications \eqref{simplicity 1} between components. As these components all commute in the bulk, this identification is not problematic. 
On the corner the flux can be viewed as the generator of the corner algebra is therefore non-commutative.  Due to this, the corner simplicity constraints are second class, and their imposition must therefore be performed with care in order to respect the Lorentz algebra structure of the boost and spin (rotation) generators.

With the simplicity constraints, we can finally express the boost and rotational parts \eqref{rotation boost} of the Gauss law as
\be
\CB^I[e]\stackrel{\Sigma}{=}\tilde{\rd}_\Gamma\tE^I+\sigma\beta(\tK\times\tE)^I\simeq0,
\q
\CR^I[e]\stackrel{\Sigma}{=}\beta\tilde{\rd}_\Gamma\tE^I-\sigma(\tK\times\tE)^I\simeq0.
\ee
One can see that in this case these two constraints are proportional to each other when $\beta^2=-1$, which is to be expected as we work in Lorentzian signature. Their imposition implies that
\be\la{ECH Gauss laws}
\tilde{\rd}_\Gamma\tE^I\simeq0,
\q
(\tK\times\tE)^I\simeq0.
\ee
In terms of $\te^I$ and the momentum aspect $\tP^I=-\sigma(\tK\times\te)^I$, this is
\be\la{realGauss}
\tilde{\rd}_\Gamma\te^I\simeq0,
\q
(\tP\times\te)^I\simeq0.
\ee
We now have at our disposal all the ingredients necessary to decompose the symplectic potentials of BF theory and ECH gravity.

\section{Decomposition of the BF symplectic potential}
\la{sec:BF potential}

The symplectic potential of BF theory is
\be
\Theta_\BF=\f{1}{2}\int_\Sigma B_{IJ}\wedge\delta\omega^{IJ}.
\ee
Using the decomposition \eqref{tensor decomposition B} in terms of boost and spin, and the fact that on $\Sigma$ only the tangential components contribute, we get
\be
\Theta_\BF=\sigma\int_\Sigma\left(\tB_I\wedge\delta\omega^{IJ}n_J-\f{1}{2}\teps_{IJK}\tS^K \wedge\delta\omega^{IJ}\right).
\ee
We now evaluate the two terms separately. For the first term, we use the variation of \eqref{decomposition of omega} to get
\be
\delta\omega^{IJ}n_J=\delta K^I+\delta\Gamma^{IJ}n_J+\sigma n^IK^J\delta n_J,
\ee
while from the variation of \eqref{covariant Gamma of n} we obtain
\be
\rd_\Gamma\delta n^I+\delta\Gamma^{IJ}n_J=0.
\ee
Together, this leads to
\be\la{delta omega n}
\delta\omega^{IJ}n_J=\delta K^I -\rd_\Gamma\delta n^I+\sigma n^IK^J\delta n_J.
\ee
The last term does not contribute when contracted with $B_I$, so we can therefore write that
\be\la{Bodom1}
B_I\wedge\delta\omega^{IJ}n_J
&=B_I\wedge(\delta K^I-\rd_\Gamma\delta n^I)\cr
&=B_I\wedge\delta K^I+\rd_\Gamma B_I\delta n^I-\rd(B_I\delta n^I).
\ee
To write the second term in the potential, let us first notice that
\be\la{Bodom2}
-\sigma\teps_{IJK}S^K\wedge\delta\omega^{IJ}=-\sigma\teps_{IJK}S^K\wedge\delta\Gamma^{IJ}+2(K\times S)_J\delta n^J,
\ee
and then use \eqref{Ss} to write
\be\la{second term in BF pot decomp}
-\sigma\teps_{IJK}S^K\wedge\delta\Gamma^{IJ}
&\stackrel{\Sigma}=(\ts\wedge\ts)_{IJ}\wedge\delta\Gamma^{IJ}\cr
&=-\ts_I\wedge\delta\Gamma^{IJ}\wedge\ts_J\cr
&=\ts_I\wedge\big(\rd_\Gamma\delta\ts^I-\delta(\rd_\Gamma\ts^I)\big)\cr
&=2\rd_\Gamma\ts_I\wedge\delta\ts^I-\delta(\ts_I\wedge\rd_\Gamma\ts^I)-\rd(\ts_I\wedge\delta\ts^I).
\ee
Noticing that \eqref{Bodom1} and \eqref{Bodom2} have two terms which combine to form the boost Gauss law $\B_I$ of \eqref{rotation boost}, and writing all the quantities on $\Sigma$ in terms of horizontal forms, we finally get
\be\la{BF decomposition}
\Theta_\BF=\Theta_\BF^\Sigma+\Theta_\BF^S,
\ee
where the bulk potential is 
\be
\Theta_\BF^\Sigma\coloneqq\int_\Sigma\left(\sigma\tB_I\wedge\delta\tK^I+\tilde{\rd}_\Gamma\ts_I\wedge\delta\ts^I+\sigma\tilde{\B}_I\delta n^I\right)-\delta\left(\f{1}{2}\int_\Sigma\ts_I\wedge\tilde{\rd}_\Gamma\ts^I\right),
\ee
and the corner one is
\be\la{BF corner potential}
\Theta_\BF^S\coloneqq-\int_S\left(\sigma\tB_I\delta n^I+\f{1}{2}\ts_I\wedge\delta\ts^I\right).
\ee
Notice that, in the bulk potential, the third term vanishes on-shell while the last one is a total variation. These two terms therefore don't contribute to the on-shell symplectic structure.

It is now a straightforward task to impose the simplicity constraints in order to obtain the gravitational potential.

\section{Decomposition of the ECH symplectic potential}
\la{sec:ECH-potential}

With the decomposition of the BF symplectic potential we can now easily obtain the decomposition of the ECH potential by plugging the simplicity constraints. We then compare this ECH potential with the GR potential of metric gravity in order to identify the relative potential, and then use this latter to study the relative charges.

\subsection{ECH symplectic potential}

Using the simplicity constraints \eqref{simplicity 2} in \eqref{BF decomposition} gives us
\be\la{ECH decomposition}
\Theta_\ECH=\Theta^\Sigma_\ECH+\Theta^S_\ECH,
\ee
with
\begin{subequations}
\be
\Theta^\Sigma_\ECH&\coloneqq\int_\Sigma\left(\sigma\tE_I\wedge\delta\tK^I+\beta\tilde{\rd}_\Gamma\te_I\wedge\delta\te^I+\sigma\tilde{\B}_I\delta n^I\right)-\delta\left(\f{\beta}{2}\int_\Sigma\te_I\wedge\tilde{\rd}_\Gamma\te^I\right),\\
\Theta^S_\ECH&\coloneqq-\int_S\left(\sigma\tE_I\delta n^I+\f{\beta}{2}\te_I\wedge\delta\te^I\right)\la{corner ECH potential}.
\ee
\end{subequations}
The ECH potential is manifestly very similar to the BF potential. The third term in the bulk vanishes on-shell of the boost Gauss constraint $\tilde{\B}_I\simeq0$, and the last one is a total variation. The key difference between the BF and ECH potentials is that the former depends on the two coframes $\tb^I$ (through $\tB^I$) and $\ts^I$, while the latter depends only on\footnote{As mentioned above, we can actually choose a notation to match $\tb^I$ with $\te^I$ from the onset.} $\te^I$. Using the form \eqref{ECH Gauss laws} of the Gauss law, we therefore get
\be
\tilde{\rd}_\Gamma\tE_I\simeq0\q\Rightarrow\q\tilde{\rd}_\Gamma\te^I\simeq0,
\ee 
meaning that the second bulk term also vanishes on-shell. While BF theory therefore possesses two bulk canonical pairs, ECH gravity only has a single bulk canonical pair. This is the usual gravitational pair expressing the fact that the extrinsic curvature $\tK^I$ is conjugated to the flux 2-form $\tE_I$. On-shell, the symplectic structure takes the simple form
\be\la{OECH}
\Omega_\ECH\simeq\sigma\int_\Sigma\delta\tE_I\wedge\delta\tK^I-\int_S\left(\sigma\delta\tE_I\delta n^I+\f{\beta}{2}\delta\te_I\wedge\delta\te^I\right).
\ee
The corner term in \eqref{OECH} is consistent with \cite{Peld_n_1994, Engle:2010kt, Bodendorfer:2013jba}. Here we have clarified how the simplicity constraints reduce the BF symplectic structure to the gravitational one.

In order to compare these results with metric gravity, it is useful to rewrite the bulk potential in terms of the momenta $\tP^I=-\sigma(\tK\times\te)^I$. This can be done using the identities (see Appendix \ref{appendix:PM} for the third one)
\be\la{EPrelations}
\tE^I=\f{1}{2}(\te\times\te)^I,
\q
\sigma\tE_I\wedge\tK^I=-\f{1}{2}\tP_I\wedge\te^I,
\q
\sigma\tE_I\wedge\delta \tK^I=\tP_I\wedge\delta\te^I-\f{1}{2}\delta(\te^I\wedge\tP_I),
\ee
leading to
\be\la{ECH with P}
\Theta_\ECH^\Sigma=\int_\Sigma\tP_I\wedge\delta\te^I+\int_\Sigma\left(\beta\tilde{\rd}_\Gamma\te_I\wedge\delta\te^I+\sigma\tilde{\B}_I\delta n^I\right)-\delta\left(\f{1}{2}\int_\Sigma\te^I\wedge\big(\tP_I+\beta\tilde{\rd}_\Gamma\te^I\big)\right),
\ee
where the Boost Gauss law now reads $\tilde{\B}_I=\big((\tilde{\rd}_\Gamma\te-\beta\tP)\times\te\big)_I$.

We now want to establish that this bulk piece coming from the tetrad gravity potential is, on-shell of the torsion, the universal piece $\Theta_\text{GR}$ common to all formulations of gravity, as stated in \cite{Freidel:2020xyx}.

\subsection{Relationship between ECH and GR potentials}

We now give the explicit proof of the relationship between the bulk ECH potential and the GR potential. Similar relationships are given in \cite{Bodendorfer_2013,Bodendorfer:2013jba,DePaoli:2018erh,Oliveri:2019gvm}, but here we establish them here in minute details. For this we have to consider the case $\sigma=-1$, but notice that we could also take $\sigma=+1$ and compare with the Gibbons--Hawking potential $\Theta_\GH$ introduced in \cite{Freidel:2020xyx}. Recall that the canonical gravitational symplectic potential is \cite{Freidel:2020xyx}
\be
\Theta_\GR=\f{1}{2}\int_\Sigma\teps(\tK\tilde{g}^{\mu\nu}-\tK^{\mu\nu})\delta\tilde{g}_{\mu\nu},
\ee
where $\tK_{\mu\nu}={\tilde{g}_\mu}^\alpha{\tilde{g}_\nu}^\beta\nabla_\alpha n_\beta$ is the extrinsic curvature tensor of the slice $\Sigma$. This potential expresses the fact that
\be 
\tP^{\mu\nu}\coloneqq\teps(\tK\tilde{g}^{\mu\nu}-\tK^{\mu\nu})
\ee
is conjugated to the induced metric $\tilde{g}_{\mu\nu}$, as we know from ADM analysis \cite{Arnowitt:1959ah,Arnowitt:1962hi}. In vacuum, this momentum satisfies the conservation law
\be
\tilde{\nabla}_\mu \tP^{\mu\nu}=0,
\ee
where $\tilde{\nabla}$ is the induced derivative on $\Sigma$. This is the vector constraint generating spatial diffeomorphisms.

This structure is of course reminiscent of the momentum aspect 2-form introduced above. When $\sigma=-1$ it is given by
\be
\tP^I\coloneqq(\tK\times\te)^I\,.
\ee
This momentum aspect was previously introduced and investigated in \cite{Freidel:2019ees,Freidel:2019ofr}. We have shown in \eqref{conservation P} that the validity of vacuum Einstein equations implies the \textit{momentum conservation}
\be
\tilde{\rd}_\Gamma \tP^I\approx0,
\ee
and in \eqref{ECH with P} that $\tP^I$ is the momentum canonically conjugated to the coframe field. As shown in Appendix \ref{appendix:PM}, for any vector-valued 1-form $M^I=M_\mu^I\rd x^\mu$ which is both tangential and horizontal we have
\be\la{momentum P identity}
\int_\Sigma \tP_I\wedge M^I=\int_\Sigma\teps\big(\tK\tilde{g}_\mu{}^\nu-\tK_\mu{}^\nu\big)M_\nu^I\te_I^\mu=\int_\Sigma{\tP_\mu}^\nu M_\nu^I\te_I^\mu.
\ee
By taking $M^I=\rd\xi^I$ one gets that the momentum conservation takes the form
\be
\tilde{\rd}_\Gamma \tP^I=(\tilde{\nabla}_\mu \tP^{\mu\nu})\te_\nu^I\approx0.
\ee 
Taking $M^I=\delta\te^I$ and using $\delta\tilde{g}_{\mu\nu}=\te^I_\mu\delta\te_{\nu I}+\te^I_\nu\delta\te_{\mu I}$ on the one hand, and $M^I=\te^I$ on the other hand, implies that 
\be
\Theta_\text{GR}=\int_\Sigma \tP_I\wedge\delta\te^I,
\q
\int_\Sigma\teps\tK=\f{1}{2} \int_\Sigma \tP_I\wedge\te^I.
\ee
We can finally state our main result, in line with \cite{Freidel:2020xyx}, which is that the ECH and GR potentials only differ by a corner symplectic potential. On-shell we have
\be\la{ECH/GR}
\boxed{\quad\Theta_\ECH\simeq\Theta_\GR+\Theta_{\ECH/\GR}-\delta\left(\int_\Sigma\teps\tK\right),\quad}
\ee
where the relative potential is the corner potential (recall that we have $\sigma=-1$)
\be\la{ThetaEC}
\Theta_{\ECH/\GR}\coloneqq\Theta^S_\ECH=\int_S\left(\tE_I\delta n^I-\f{\beta}{2}\te_I\wedge\delta\te^I\right).
\ee
This can be put in parallel with the decomposition of the Einstein--Hilbert potential performed in \cite{Freidel:2020xyx} (see (2.14) there), which is
\be\la{EH/GR1}
\Theta_\EH=\Theta_\GR+\Theta_{\EH/\GR}-\delta\left(\int_\Sigma\teps\tK\right).
\ee
Moreover, continuing this comparison with the metric case, let us point out as a remark that we can express the relation \eqref{ECH/GR} in terms of the boundary Lagrangian\footnote{This form of Lagrangian has been used to write  the covariant Gibbons--Hawking term \cite{0264-9381-4-5-011,Wieland_2011,Bodendorfer:2013hla} plus a Holst boundary term vanishing on-shell of the torsion \cite{Freidel:2016bxd}. We have written it here for an arbitrary $\sigma$.}
\be
L_{\ECH/\GR}\coloneqq\f{1}{2}\Big(\sigma*(e\wedge e)_{IJ}\wedge\rd_\omega n^In^J-\beta e_I\wedge\rd_\omega e^I\Big)
\ee
as
\be\la{variation ECH/GR Lagrangian}
\delta L_{\ECH/\GR}+\rd\theta_{\ECH/\GR}\stackrel{S}{\simeq}\theta_\ECH-\theta_\GR.
\ee
This is the exact analog of equation (2.15) in \cite{Freidel:2020xyx}, which expresses the same relationship between the EH and GR metric formulations. The proof of this relation is given in Appendix \ref{AppD}. This formula means that the first order GR Lagrangian is given by $L_\GR=L_\ECH-\rd L_{\ECH/\GR}$. We will come back to this boundary Lagrangian when studying the whole boundary dynamics in future work.

Taking the variation of \eqref{EH/GR1} to obtain the symplectic structures we have
\be\la{ECH GR Omegas}
\Omega_\ECH\simeq\Omega_\GR+\Omega_{\ECH/\GR},
\ee
with
\be\la{EHsympdec}
\Omega_\GR=\int_\Sigma\delta \tP_I\wedge\delta\te^I,
\q
\Omega_{\ECH/\GR}=\int_S\left(\delta \tE_I\wedge\delta n^I-\f{\beta}{2}\delta\te_I\wedge\delta\te^I\right).
\ee

We can now use the relative symplectic structure in \eqref{EHsympdec} to discuss the relative charges. Before going on to this, let us discuss for completeness the relationship between the ECH and Einstein--Hilbert potentials.

\subsection{Relationship between ECH and EH potentials}

In our previous work \cite{Freidel:2020xyx} we have established the relationship between the canonical GR potential and the Einstein--Hilbert potential. Similarly, in the previous section we have established the relationship between the GR and the Einstein--Cartan--Holst potential. This is summarized in the two identities
\begin{subequations}
\be
\Theta_\EH&=\Theta_\GR+\Theta_{\EH/\GR}-\delta\left(\int_\Sigma\teps\tK\right),\\
\Theta_\ECH&\simeq\Theta_\GR+\Theta_{\ECH/\GR}-\delta\left(\int_\Sigma\teps\tK\right),
\ee
\end{subequations}
where $\simeq$ means that the torsion equation of motion has been used, and where the corner symplectic potentials are (we continue to restrict our analysis to the case where $n^I$ is time-like, so $\sigma=-1$) \cite{Freidel:2020xyx}
\be\la{TECH}
\Theta_{\EH/\GR}=\f{1}{2}\int_S\sqrt{q}\,s_\mu\delta n^\mu,
\q
\Theta_{\ECH/\GR}=\int_S\left(\tE_I\delta n^I-\f{\beta}{2}\te_I\wedge\delta\te^I\right).
\ee
Here $\hat{s}=s^\mu\partial_\mu$ denotes\footnote{This vector is denoted by $\hat{\ts}=\ts^\mu\partial_\mu$ in our previous paper \cite{Freidel:2020xyx}, where we consider a more generic geometrical setup at the corner with two pairs of normals: $(n^\mu,\tilde{s}^\mu)$ is used when the slice is space-like, and $(s^\mu,\bar{n}^\mu)$ when it is time-like, with a boost angle between $\hat{n}$ and $\hat{s}$. When this boost angle vanishes we have $\hat{s}=\hat{\tilde{s}}$. Here we work on a space-like slice, but drop the tilde in order to have lighter notations in this section.} the outward unit vector tangent to the slice $\Sigma$ and normal to $S$. By taking the difference of the corner symplectic potentials, we get a direct relationship between Einstein--Hilbert and Einstein--Cartan--Holst potentials, namely
\be\la{ECH/EH}
\Theta_\ECH-\Theta_\EH\coloneqq\Theta_{\ECH/\EH}=\Theta_{\ECH/\GR}-\Theta_{\EH/\GR}.
\ee 
This gives the relative potential between the ECH and EH formulations, and enables to study the relationship between the diffeomorphism Komar charge $\CH_\EH[\xi]$ and the ECH diffeomorphism charge $\CH_\ECH[\xi]$. Let us expand a bit on this formula and explain how it is related to the work \cite{DePaoli:2018erh}.

Since all quantities are pulled back to $S$, which is of codimension-2, it is convenient to introduce the pull-back on $S$ of the coframe. We focus again on the time-like $\hat{n}$ case, so that $\sigma=-1$, and use the availability of the space-like unit vector $\hat{s}$ normal to the surface $S$. Given these two normal vectors such that $g^{\mu\nu}n_\mu n_\nu=-1$, $g^{\mu\nu}s_\mu s_\nu=1$, and $g^{\mu\nu}n_\mu s_\nu=0$, we introduce the induced coframes $\te^I\coloneqq e^I+\un n^I$, $\bar{e}^I\coloneqq e^I-\us s^I$, and $\bte^I=\tbe^I\coloneqq e^I+\un n^I-\us s^I$. This last coframe is the pull-back to $S$ of the bulk coframe field, i.e. we have
\be
e^I\stackrel{\Sigma}{=}\te^I\stackrel{S}=\bte^I.
\ee
These coframes are tangential in the sense that $\bar{e}^Is_I=0=\te^In_I$, and satisfy $\te^I=\bte^I+\us s^I$ and $\bar{e}^I=\tbe^I-\un n^I$.

In \cite{DePaoli:2018erh}, De Paoli and Speziale have studied the relationship between the Einstein--Cartan--Holst and Einstein--Hilbert formulations of gravity. They have introduced a corner potential for tetrad gravity such that internal gauge invariance is restored, i.e. the Lorentz charges on the corner vanish, and such that the Komar expression is recovered for Hamiltonian charge associated with tangent diffeomorphisms. In order to write this corner term we use the inverse $\hat{e}_I=e_I^\mu\partial_\mu$ to introduce the variational 1-form
\be
\varpi^{IJ}\coloneqq\hat{e}^{[I}\ip\delta e^{J]}.
\ee
The De Paoli--Speziale (DPS) corner term\footnote{Please note that we have adjusted signs and numerical factors with respect to \cite{DePaoli:2018erh} in order to fit our conventions.} is
\be\la{DPS}
\Theta_\DPS\coloneqq\f{1}{2}\int_S\varpi_{IJ}E^{IJ}.
\ee
We are now going to show this DPS corner term is related to the various potentials and relative potentials introduced so far. This is the series of equalities
\be\la{DPS relative equalities}
\Theta_\DPS=\Theta_{\ECH/\GR}-\Theta_{\EH/\GR}=\Theta_\ECH-\Theta_\EH=\Theta_{\ECH/\EH}.
\ee
Introducing $\btE_I=(\bte\times\bte)_I$, we show in Appendix \ref{app:DPS} that
\be\la{DPS identity}
*(e\wedge e)_{IJ}\varpi^{IJ}\stackrel{S}{=}2\btE_I\delta n^I-\sqrt{q}\,s_\mu\delta n^\mu.
\ee
For the Holst term we have
\be
(\te\wedge\te)_{IJ}\varpi^{IJ}=(\te\wedge\te)_{IJ}\hat{e}^I\ip\delta e^J=-\te_I\wedge\delta\te^I.
\ee
Putting this together shows the desired result, namely that
\be
\Theta_\DPS=\f{1}{2}\int_S\varpi_{IJ}E^{IJ}=\int_S\left(\tE_I\delta n^I-\f{\beta}{2}\te_I\wedge\delta\te^I-\f{1}{2}\sqrt{q}\,s_\mu\delta n^\mu\right)=\Theta_{\ECH/\EH}.
\ee
This identity clearly shows that the DPS corner symplectic potential contains three canonical pairs, namely $(\tE,n^I)$, $(\te_1^I,\te_2^I)$, and $(\sqrt{q}\,s_\mu,n^\mu)$.

\subsection{Relative charges}

Let us now go back to the relationship between the two formulations GR and ECH. Since we have two different expressions for the gravitational potentials, we also have two different expressions for the corner charges of symmetry associated with Lorentz transformations and diffeomorphisms. This is captured by the notion of relative charge defined as
\be\la{relative ECH/GR charge}
\CH^S_{\ECH/\GR}=\CH^S_\ECH-\CH^S_\GR.
\ee
The charges $\CH^S_\ECH$ and $\CH^S_\GR$ are evidently expressed as boundary integrals after imposing the bulk equations of motion \cite{Freidel:2020xyx}. The relative charge can then be expressed either as the difference $\CH^S_\ECH-\CH^S_\GR$, or directly as a canonical charge coming from the relative symplectic potential. This last option amounts to treating the relative charge and the relative potential (and symplectic structure) as standing on their own feet. Let us elucidate this point of view by analyzing these relative charges.

We start with the easier case of the charges for Lorentz transformations with parameter $\alpha^{IJ}$. In this case, the GR charge vanishes identically, so the relative charge $\CH^S_{\ECH/\GR}[\alpha]$ is simply equal to $\CH^S_\ECH[\alpha]$ in \eqref{ECH charges} and given by
\be\la{relative Lorentz}
\CH^S_{\ECH/\GR}[\alpha]=\CH^S_\ECH[\alpha]=\f{1}{2}\int_{S}\alpha_{IJ}E^{IJ}=-\int_S\left(\alpha^I_\perp\tE_I-\f{\beta}{2}\alpha_{IJ}(\te\wedge\te)^{IJ}\right),
\ee
where we have used the first equality in \eqref{E on Sigma} with $\sigma=-1$. Using the explicit expression \eqref{EHsympdec} for $\Omega_{\ECH/\GR}$ and the transformation rule $\delta_\alpha V^I=-\alpha^I{}_JV^J$ for $V^I=(\tE^I,n^I,\te^I)$, it is easy to show that the relative charge is in fact a canonical charge for the corner symplectic structure, i.e. that
\be
-\delta_\alpha\ip\Omega_{\ECH/\GR}=\delta\CH^S_{\ECH/\GR}[\alpha].
\ee 

We now prove the same results for the diffeomorphism charges. The GR and ECH charges for diffeomorphisms tangent to the corner are simply obtained by the on-shell Noether expression $\CH^S[\xi]\approx{\CL_\xi}\ip\Theta$ and accordingly they are given by\footnote{On-shell of all the torsion-free equations one can replace $\omega$ by the spin connection $\gamma[e]$. It is however sometimes also useful to keep $\omega$ and first decompose it as \eqref{decomposition of omega} before going on-shell.}
\be
\CH^S_\GR[\xi]\approx\int_S\xi\ip\te^I\tP_I,
\q
\CH^S_\ECH[\xi]\approx\f{1}{2}\int_S\xi\ip\omega_{IJ}E^{IJ}.
\ee
We present in Appendix \ref{appendix:relative diffeo} two computations of the relative charge. One can directly compute the subtraction $\CH^S_{\ECH/\GR}[\xi]=\CH^S_\ECH[\xi]-\CH^S_\GR[\xi]$, as done in \ref{appendix:relative diffeo 1} (where as a byproduct we also decompose the diffeomorphism charge of BF theory). Alternatively, one can compute the relative charge as the canonical charge of the relative symplectic structure. From the expression \eqref{TECH} of the relative potential, it is immediate to see that the contraction $\CL_\xi\ip\Theta_{\ECH/\GR}$,
 for $\sigma=-1$, yields the relative charge 
\be\la{relative diffeo charge}
\CH^S_{\ECH/\GR}[\xi]=\int_S\left(\tE_I\CL_\xi n^I-\f{\beta}{2}\te_I\wedge\CL_\xi\te^I\right)\,,
\ee
in agreement with \eqref{H-ECH/GR}.

For the sake of completeness, we give in Appendix \ref{appendix:relative diffeo 2} yet another proof of the relationship \eqref{relative ECH/GR charge} between the diffeomorphism charges. There we show how the GR charge written as $\CH^S_\ECH[\xi]-\CH^S_{\ECH/\GR}[\xi]$ and in the parametrization introduced in Section \ref{sec:corner} contains two terms. One is a  Holst piece which yields the so-called topological Komar charge (see also \cite{DePaoli:2018erh}) and it vanishes on-shell. The second one is a gravitational piece which yields   the Brown--York charge expressed in first order variables, as expected for the GR charge \cite{Freidel:2020xyx}.


Finally, we can also compare the diffeomorphism ECH charge with the EH Komar charge. The difference is encoded in the relative charge $\CH^S_{\ECH/\EH}[\xi]=\CH^S_\ECH[\xi]-\CH^S_\EH[\xi]$. To compute this relative charge directly, let us first recall that on-shell of the torsion equations the Lorentz connection $\omega^{IJ}$ becomes the compatible torsion-less connection
\be
\gamma_\mu^{IJ}[e]=\left(\delta_\mu^\alpha\big(e^{I\beta }\delta^J_K-e^{J\beta }\delta^I_K\big)-e^{I\alpha}e^{J\beta }e_{\mu K}\right)\partial_{[\alpha}e_{\beta]}{}^K.
\ee
Using this we get that
\be
\xi\ip\gamma^{IJ}=\hat{e}^{[I}\ip\CL_\xi e^{J]}-\hat{e}^I\ip\hat{e}^J\ip\rd\underline{\xi},
\ee
where $\underline{\xi}=\xi_\mu\rd x^\mu$. This enables us to decompose the diffeomorphism charge as
\be
\CH^S_\ECH[\xi]=\f{1}{2}\int_S\xi\ip\gamma_{IJ}E^{IJ}=\f{1}{2}\int_S E^{IJ}(\hat{e}_{I}\ip\CL_\xi e_{J})-\f{1}{2}\int_S E^{IJ} \hat{e}_I\ip\hat{e}_J\ip\rd\underline\xi.
\ee
The last term on the right-hand side is precisely the Komar charge, telling us that
\be
\CH^S_{\ECH/\EH}[\xi]=\CH^S_\ECH[\xi]-\CH^S_\EH[\xi]=\f{1}{2}\int_S E^{IJ}(\hat{e}_{I}\ip\CL_\xi e_{J}),
\ee
which immediately shows that this is also the canonical charge derived from the relative potential \eqref{DPS relative equalities}.

\section{Normal coframe, corner simplicity constraints, and corner metric}
\la{sec:corner}

So far we have studied in details the relationship between the ECH, GR, and EH potentials and charges. It is now time to focus exclusively on the corner symplectic potential $\Theta^S_\ECH$ \eqref{corner ECH potential}. After pointing out why the corner phase space  has to be handled with care, we show that it contains important physical information, in particular the symplectic structure of the tangential corner metric and its algebra. The corner potential also guides us towards the corner simplicity constraints which we study in details in \cite{Edge-Mode-III}. At the end of this section, we will be in position to make the last conceptual jump towards edge modes, by promoting the corner fields to new independent degrees of freedom, and shifting once and for all the focus to the corner symplectic structure.

\subsection{Danger around the corner}
\la{sec:danger}

In Section \ref{sec:simplicity} we put the reader on alert about the danger of using the naive pull-back of the bulk simplicity constraints \eqref{simplicity 1} to the corner. Let us explicitly show on the example of the Lorentz charges where such a wrong turn would lead us. First, let us recall that in BF theory the $B$ field conjugated to $\omega$ is a commutative bulk variable, as can be seen on \eqref{bare BF potential}. However, after imposing the Gauss law the $B$ field at the corner becomes non-commutative. This is evidently also the case in ECH gravity with the simple field $B^{IJ}=E^{IJ}[e]$ \cite{Cattaneo:2016zsq}. This simply follows from the identity
\be 
\CH^S_\ECH[\alpha]=\f{1}{2}\int_{S}\alpha_{IJ}E^{IJ}\simeq\f{1}{2}\int_\Sigma\rd_A\alpha_{IJ}\wedge E^{IJ}
\ee
and the Poisson bracket $\{\CH^S_\ECH[\alpha],\CH^S_\ECH[\alpha]\}\simeq\CH^S_\ECH[[\alpha,\beta]]$.

Now, notice that a naive imposition on the corner of the simplicity constraints in \eqref{relative Lorentz} would yield the Lorentz charge
\be
\CH^S_\ECH[\alpha]=\f{1}{2}\int_{S}\alpha_{IJ}E^{IJ}\stackrel{\dangersign}=\sigma\int_S\tE_I\big(\alpha_\per^I-\beta\alpha_{\para}^I\big).
\ee
One would therefore be tempted to call $\tE_I$ the boost generator and $\beta\tE_I$ the rotation generator. This identification is however clearly at odds with the Poisson brackets of the boost and spin generators $\tB^I$ and $\tS^I$ which one can read off from the BF corner symplectic potential \eqref{BF corner potential} before imposing the simplicity constraints (this phase space structure will be analyzed in detail in \cite{Edge-Mode-III}).  In the bulk of $\Sigma$ the identification $\tS^I\stackrel{\Sigma}=\beta\tB^I$ is harmless since all the components of $B^{IJ}$ commute with each other (as well as the components of $E^{IJ}$). At the corner $\tB^I$ is the canonical generator for boosts while $\tS^I$ is the canonical generator of rotations, so the
identification $\tS^I\stackrel{S}=\beta\tB^I$ cannot be done without harm to the Poisson structure\footnote{The fact that boundary continuity equations should be treated as second class constraints was first realized by Smolin \cite{Smolin:1998qp}}.
At the quantum level we also cannot naively identified the rotation and boost operators.

This means that we cannot simply understand the corner simplicity constraint as arising by continuity from the bulk one. That is why one needs to be very careful in identifying  the flux (and eventually the area) with the Lorentz spin generator when talking about the corner phase space. We will see in \cite{Edge-Mode-III} how this apparently innocent fact is actually at the core of the persisting confusion about the reconciliation of fundamental discreteness of quantum geometry and Lorentz invariance. In fact, such a reconciliation follows straightforwardly from the reconciliation between the bulk and the corner phase space structures. This is where the edge modes come to the rescue, as this is exactly their main role. We elaborate extensively on this crucial aspect in Section \ref{sec:edge modes}.

For the time being, as a warm-up to the introduction of the edge modes, let us present an alternative way of lifting this difficulty of reconciling bulk simplicity with corner non-commutativity. This is provided by performing a more ``symmetric'' rewriting of $\tE^{IJ}$ so that the simplicity 
can be implemented straightforwardly in the corner phase space while intertwining the Lorentz algebra (at the cost of losing for the moment the clear geometrical split between boosts and rotations).

\subsection{Normal coframe parametrization}
\la{sec:boundary with nIJ}

The corner phase space can be elegantly described by using the so-called normal coframe parametrization. Let us consider the Lie algebra-valued horizontal 1-form $\n^{IJ}$ defined as
\be\la{nmode}
\n^{IJ}\coloneqq\sigma\hat{n}\ip(e\wedge e)^{IJ}=2\sigma n^{[I}\te^{J]}.
\ee
From this definition, one can see that the normal component (up to a sign) of $\n^{IJ}$ is the horizontal coframe, i.e. $n_I\n^{IJ}=\te^J$. This object satisfies many relations which are proven in Appendix \ref{appendix:En}. In particular, it enables us to write the horizontal part \eqref{E on Sigma} of the gravitational 2-form as
\be\la{E in terms of n}
\tE^{IJ}=-\f{\sigma}{4}(*+\beta)[\n\wedge\n]^{IJ}.
\ee
This normal coframe also provides an elegant alternative parametrization of the corner symplectic potential. In order to see this, we use the fact that the Einstein--Cartan piece of the corner potential $\Theta^S_\ECH$ can be rewritten as
\be\la{E=n^2}
\tE_I\delta n^I=\f{1}{2}{\eps}_{IJKL}n^J\te^K\wedge\delta(n^I\te^L)=\f{1}{8}{\eps}_{IJKL}\n^{IJ}\wedge\delta\n^{KL}=\f{1}{4}(*\n)_{IJ}\wedge\delta\n^{IJ},
\ee
while the Holst component can be written as 
\be
\f{1}{2}\te_I\wedge\delta\te^I=\f{\sigma}{2}(n_I\te_J-n_J\te_I)\wedge(\delta n^I\te^J+n^I\delta\te^J)=\f{\sigma}{4}\n_{IJ}\wedge\delta\n^{IJ}.
\ee
With this we get
\be\la{TS2}
\boxed{\quad\Theta_\ECH^S=-\f{\sigma}{4}\int_S(*+\beta)\n_{IJ}\wedge\delta\n^{IJ},\quad}
\ee
and the corner symplectic structure is therefore
\be\la{Omega-nn}
\Omega_\ECH^S=-\f{\sigma}{4}\int_S(*+\beta)\delta\n_{IJ}\wedge\delta \n^{IJ}.
\ee
Note that this parametrization of the corner symplectic structure is used in Appendix \ref{appendix:relative diffeo 2} as a way to compute the relative diffeomorphism charge. This can be seen as an interesting consistency check for our formulas, and as an exercise in manipulating the corner variables $\n^{IJ}$ which we have introduced.

\subsection{Corner simplicity constraints}
\la{sec:boundary simplicity}

It is important at this point to stop and think about the number of independent variables in the various parametrizations of the corner phase space. For this, let us forget momentarily about relation \eqref{nmode}. The unconstrained corner phase space with canonical variables $\n^{IJ}_a$, with $a=1,2$ an index tangent to $S$,  is then 12-dimensional. This amounts to working with the BF corner potential \eqref{BF corner potential} in terms of variables $(\tB^I,n^I,\te^I_a)$, which because of the relations $\tB^In_I=0=\te^I_an_I$ and $n^2=\sigma$ gives indeed a 12-dimensional phase space\footnote{In \cite{Edge-Mode-III} we will even relax these 4 kinematical constraints and start from a 16-dimensional corner phase space. This will enable to unravel an interesting symmetry breaking pattern.}. Another convenient parametrization is to use the $3+3+6=12$ objects
\be\la{s q J def}
s_{ab}\coloneqq*\n^{IJ}_{(a}\n_{b)IJ},
\q
q_{ab}\coloneqq\f{\sigma}{2}\n^{IJ}_{(a}\n_{b)IJ},
\q
J_{IJ}\coloneqq-\f{\sigma}{4}(*+\beta)[\n\wedge\n]_{IJ}.
\ee
To go back to the gravitational phase space, one should impose the 3 corner simplicity constraints $\tB^I=*(\te\wedge\te)^{IJ}n_J$. These can elegantly be imposed directly on $\n^{IJ}$ as the conditions
\be\la{n simplicity}
s_{ab}=0,
\ee
which indeed imply that there exists a vector $n^I$ (which can always be chosen to be normalized) and a form $\te^I_a$ such that $\n_a^{IJ}=2\sigma n^{[I}\te_a^{J]}$. As we will explain in details in \cite{Edge-Mode-III}, the 3 simplicity constraints contain actually a first class constraint and 2 second class constraints, and as such remove 4 degrees of freedom. The reduced phase space is then 8-dimensional\footnote{The corner metric $q_{ab}$ and the Lorentz generator $J_{IJ}$ satisfy the scalar relation $\big((*+\beta)^{-1}J\big)^2\propto\det(q)$, which is why they together describe 8 variables and not 9.}, and parametrized by the components of $q_{ab}$ and $J_{IJ}$. When the simplicity constraints are satisfied $q_{ab}=\te^I_a\te^J_b\eta_{IJ}$ is the corner metric, while $J_{IJ}$ then coincides with the gravitational 2-form \eqref{E in terms of n}. We will now compute Poisson brackets and show that $J_{IJ}$ is the generator of corner Lorentz transformations.\footnote{$J_{IJ}$ is just a rewriting at the corner of $B_{IJ}$ before simplicity, and of the gravitational 2-form $E_{IJ}$ after simplicity.}

Now that we have motivated the use of the variables $\n^{IJ}$ to parametrize the phase space as \eqref{Omega-nn}, we can derive the associated Poisson brackets and use them to study the algebra of the objects \eqref{s q J def}.

\subsection{Poisson brackets}

Let us now invert the symplectic form \eqref{Omega-nn} in order to derive the corner phase space Poisson brackets. Given a Lie algebra-valued 1-form $\lambda^{IJ}$, the transformation $\delta_\lambda\n^{IJ}=\lambda^{IJ}$ is Hamiltonian, i.e. satisfies $\delta_\lambda\ipp\Omega_\ECH^S=-\delta\mathsf{N}(\lambda)$, where the corresponding Hamiltonian generator is
\be
\NN(\lambda)=\f{\sigma}{2}\int_S(*+\beta)\lambda_{IJ}\wedge\n^{IJ}.
\ee
The Poisson bracket of these Hamiltonians can be computed in two equivalent ways. First, using the covariant phase space formalism, it is given by
\be
\{\NN(\lambda),\NN(\lambda')\}=\delta_\lambda\NN(\lambda')=-\f{\sigma}{2}\int_S(*+\beta)\lambda_{IJ}\wedge\lambda'^{IJ}.
\ee  
On the other hand, we can write it in terms of the fundamental Poisson bracket which we are trying to compute as
\be
\{\NN(\lambda),\NN(\lambda')\}=\f{1}{4}\iint_S(*+\beta)\lambda_{aIJ}(x)(*+\beta)\lambda'_{cKL}(y)\{\n_b^{IJ}(x),\n_d^{KL}(y)\}\eps^{ab}\eps^{cd}\rd^2x\rd^2y.
\ee
Making the ansatz
\be
\{\n_b^{IJ}(x),\n_d^{KL}(y)\}=\eps_{bd}\delta^{(2)}(x,y)T^{IJKL},
\ee
where $T^{IJKL}$ is a tensor to be determined, and comparing these two expressions for the Poisson bracket, we arrive at the condition
\be
\lambda_{IJ}={-}\f{\sigma}{2}(*+\beta)\lambda^{KL} T_{IJKL}.
\ee
Using the identity
\be
\sigma(*+\beta)(*-\beta)=1-\sigma\beta^2,
\ee
we get the Poisson structure
\be\la{n bracket}
\{\n_a^{IJ}(x),\n_b^{KL}(y)\}=\f{1}{1-\sigma\beta^2}\eps_{ab}\left(\beta(\eta^{IK}\eta^{JL}-\eta^{IL}\eta^{JK}) - \eps^{IJKL} \right)\delta^{(2)}(x,y).
\ee
Here $\eps_{ab}=\eps^{ab}$ is the totally skew symbol with $\eps_{12}=+1$. It enters the wedge product of forms as $\alpha\wedge\beta=(\alpha_a\eps^{ab}\beta_b) \rd^2x$. Finally, $\delta^{(2)}$ denotes the Dirac distribution normalized by $f(x)=\int_S\delta^{(2)}(x,y)f(y) \rd^2 y$.

\subsection{Corner metric and its algebra}
\la{sec:bmetric}

Now that we have the Poisson brackets arising from the symplectic form \eqref{Omega-nn}, we can evaluate the algebra of the simplicity constraint generators $s_{ab}$, of the metric components $q_{ab}$, and of the Lorentz generators $J_{IJ}$ given in \eqref{s q J def}. Note that these calculations use only the brackets \eqref{n bracket} and the definitions \eqref{s q J def}, but not explicitly the relation \eqref{nmode} solving the simplicity constraints (and giving to $q_{ab}$ its interpretation as the metric).

A straightforward calculation by means of the bracket \eqref{n bracket} reveals the second class nature of the simplicity constraints. Explicitly, one finds
\be
\{s_{ab},s_{cd}\}
&=4*\n^{IJ}_{(a}*\n^{KL}_{(c}\{\n_{b)IJ},\n_{d)KL}\}\cr
&=-\f{4}{1-\sigma\beta^2}*\n^{IJ}_{(a}*\n^{KL}_{(c}\eps_{b)d)}(\eps_{IJKL}-2\beta\eta_{KI}\eta_{JL})\cr
&=-\f{2\sigma}{1-\sigma\beta^2}\left(s_{ac}\eps_{bd }+s_{bc}\eps_{ad}+s_{ad}\eps_{bc}+s_{bd}\eps_{ac}\right)\cr
&\phantom{=\ }+\f{4\beta}{1-\sigma\beta^2}\left(q_{ac}\eps_{bd}+q_{bc}\eps_{ad}+q_{ad}\eps_{bc}+q_{bd}\eps_{ac}\right).
\ee
This shows that 2 of these constraints are second class, while 1 is first class. We provide the explicit separation into first and second class components in \cite{Edge-Mode-III}, where we study in details the imposition of the simplicity constraints both at the classical and quantum level. These details are not necessary at this point.

Then, one can check that the Lorentz generators $J_{IJ}$ are Dirac observables with respect to the simplicity constraints as $\{s_{ab},J_{IJ}\}=0$, and that they satisfy as expected the $\sll(2,\mathbb{C})^S$ Lie algebra commutation relations
\be\la{corner Lorentz algebra}
\{J_{IJ}(x),J_{KL}(y)\}=\big(\eta_{JK}J_{IL}+\eta_{IL}J_{JK}-\eta_{IK}J_{JL}-\eta_{JL}J_{IK}\big)(x)\delta^2(x,y).
\ee

Focusing on the corner metric, one first notices that the 3 components defined in \eqref{s q J def} are not Dirac observables, as their bracket with the simplicity constraints is
\be
\{q_{ab}(x),s_{cd}(y)\}
&=\f{\sigma\beta}{1-\sigma\beta^2}\big(s_{ac}\eps_{bd}+s_{bc}\eps_{ad}+s_{ad}\eps_{bc}+s_{bd}\eps_{ac}\big)(x)\delta^2(x,y)\cr
&\phantom{=\ }-\f{2\sigma}{1-\sigma\beta^2}\big(q_{ac}\eps_{bd}+q_{bc}\eps_{ad}+q_{ad}\eps_{bc}+q_{bd}\eps_{ac}\big)(x)\delta^2(x,y).
 \ee
However, defining the components
\be
q'_{ab}\coloneqq q_{ab}+\f{\sigma}{2\beta}s_{ab}
\ee
provides Dirac observables. Indeed, the bracket is found to be
\be\la{q' bracket}
\{q'_{ab}(x),s_{cd}(y)\}=-\f{1}{\beta}\big(s_{ac}\eps_{bd}+s_{bc}\eps_{ad}+s_{ad}\eps_{bc}+s_{bd}\eps_{ac}\big)(x)\delta^2(x,y)\approx0,
\ee
where the symbol $\approx$ means that we have imposed the corner simplicity constraints $s_{ab}\approx 0$. Therefore, while $q'_{ab}\approx q_{ab}$, we need to use the expression \eqref{q' bracket} in order to correctly compute the algebra of the remaining corner charges without having to introduce the Dirac bracket. Explicitly, the remaining bracket is that of the metric with itself, and is given by
\be\la{qqalg}
\boxed{\quad\{q'_{ab}(x),q'_{cd}(y)\}=-\f{1}{\beta}\big(q'_{ac}\eps_{bd }+q'_{bc}\eps_{ad }+q'_{ad}\eps_{bc}+q'_{bd}\eps_{ac}\big)(x)\delta^2(x,y).\quad}
\ee

This calculation shows that the corner metric satisfies an $\sll(2,\mathbb{R})^S$ Lie algebra, in line with the analysis of \cite{Freidel:2015gpa,Freidel:2018pvm}. Importantly, it also reveals that the non-commutativity of the corner metric components is due to the presence of a non-vanishing Barbero--Immirzi parameter $\gamma=\beta^{-1}$. Taking the limit $\beta\rightarrow\infty$ shows that in  metric gravity, the corner metric components commute with each other. This highlights the role of the Barbero--Immirzi parameter in the context of gravity. The usual viewpoint is that a finite value of $\beta$ does not affect the classical bulk theory. However, we see here that it does have important implications on the corner and for its symmetry algebra: It gives rise to non-vanishing rotation charges in \eqref{relative Lorentz}, and yields an extra $\sll(2,\mathbb{R})$ factor in the corner symmetry algebra.

In summary, the Lorentz charges and the corner metric components form together an $\sll(2,\mathbb{C})\oplus\sll(2,\mathbb{R})_\para$ algebra. This gives a priori 9 generators, but these are constrained by a relation between the two $\sll(2,\mathbb{C})$ Casimirs and the $\sll(2,\mathbb{R})_\para$ Casimir associated to the corner area. This condition is a generalization of the Casimir balance equation derived in \cite{Freidel:2015gpa,Freidel:2018pvm,Freidel:2019ees} once the restriction of the time gauge is lifted. Taking into account this balance equation, we therefore recover 8 physical corner degrees of freedom. All this will be studied in great details in the follow-up paper \cite{Edge-Mode-III}.

\subsection{A glimpse into the quantization and discreteness of area}

We now come to the final main result of the paper. Having access to a non-commutative corner metric, we can investigate the fate of its quantum spectrum. In order to do so we need to introduce some regularization.

Let us assume that we choose a kinematical reference metric $q_0$ on $S$, and furthermore that $D\subset S$ is an infinitesimal (with respect to $q_0$) disk inside $S$ with $A_0(D)\coloneqq\int_D\sqrt{q_0}\,\rd^2x$. We can then smear the local charge generators along $D$ and define 
\be
A(D)\coloneqq\int_D\sqrt{q}\,\rd^2x,
\q
Q_{ab}(D)\coloneqq\int_Dq_{ab}\sqrt{q_0}\,\rd^2x.
\ee
These observables are such that $\det Q(D)/A(D)^2 \to \det{q_0}$ in the limit where $ A_0(D)\to 0$. 

Since the original algebra \eqref{qqalg} is ultra-local, we have that 
\be
\{Q_{ab}(D),Q_{cd}(D')\}=0,\quad\mathrm{ if}\q D\cap D'=\emptyset.
\ee The only non-trivial commutation relation happens on each infinitesimal disk. In order to evaluate this algebra we need to choose coordinates on $D$ such that $q_{0ab}(x)$ is constant on $D$ and such that $\sqrt{q}_0=1$. This is always possible, and we assume that this choice is made. With this choice we get from \eqref{qqalg} that $\{A(D),Q_{ab}(D)\}=0$, while 
\be
\{Q_{ab}(D),Q_{cd}(D)\}=-\f{1}{\beta}\big(Q_{ac}(D)\eps_{bd}+Q_{bc}(D)\eps_{ad}+Q_{ad}(D)\eps_{bc}+Q_{bd}(D)\eps_{ac}\big).
\ee
It is convenient to then define the new generators
\be
J_0\coloneqq\f{\beta}{4}\big(Q_{11}(D)+Q_{22}(D)\big),
\q
J_1\coloneqq\f{\beta}{4}\big(Q_{11}(D)-Q_{22}(D)\big),
\q
J_2\coloneqq\f{\beta}{2}Q_{12}(D).
\ee
The quantum algebra $[\cdot,\cdot]=-i \{\cdot,\cdot\}$ is then simply the $\sll(2,\mathbb{R})$ algebra with commutators\footnote{As an intermediate step, one has
\be
\{Q_{11}(D),Q_{22}(D)\}=\f{4}{\beta}Q_{12}(D),
\q
\{Q_{11}(D),Q_{12}(D)\}=\f{2}{\beta}Q_{11}(D),
\q
\{Q_{22}(D),Q_{12}(D)\}=-\f{2}{\beta}Q_{22}(D).
\ee}
\be
[J_0,J_1]=-iJ_2,
\q
[J_0,J_2]=iJ_1,
\q
[J_1,J_2]=iJ_0,
\ee
with $J_0$ the elliptic generator. The Casimir is then found to be
\be
C=J_0^2-J_1^2-J_2^2=\f{\beta^2}{4}\det Q(D).
\ee
In a unitary representation the Casimir is real and labelled by a weight $\lambda$ where
\be
C=\lambda(\lambda-1)=\left(\lambda-\f{1}{2}\right)^2-\f{1}{4}.
\ee
A priori, the Casimir can be positive for the discrete series of representations where $\lambda\in\mathbb{Z}$, and negative for the continuous series where $\lambda=\f{1}{2}+is$. The operator $\det Q(D)$ has to be a positive operator if we want the metric $q_{ab}$ to be the metric of a 2-dimensional space-like surface. The condition $\det Q(D)>0$ means that $C>0$, and therefore that the surface area spectrum is labelled by the discrete representations. Since in the limit of infinitesimal disk one has that $A(D)\simeq\sqrt{\det{Q(D)}}$, one concludes that the area spectrum also has to be discrete.

Note that this discretization and regularization is still a bit naive. It gives us just a hint of the quantization of the area spectrum. A more appropriate treatment that respects the diffeomorphism symmetry on the sphere needs to be given \cite{DFMS} (see also \cite{Freidel:2016bxd,Freidel:2019ees, Wieland:2017cmf}).

\section{Turning on the edge modes}
\la{sec:edge modes}

We have now shown that the symplectic potential of ECH gravity differs from the canonical GR potential by a corner term. This is the content of our claim that all formulations of gravity share the same bulk symplectic potential and differ only by a corner potential. It means that although the GR formulation of gravity is equivalent to tetrad gravity in the bulk, it is not equivalent as a theory including boundaries. Tetrad gravity has additional charges associated with local Lorentz transformations at the corner, and therefore a bigger corner symmetry algebra\footnote{When comparing ECH gravity with EH gravity one cannot fairly say which theory has a bigger corner symmetry algebra, since the former has Lorentz charges while the latter surface boost charges.}. We have also shown that the corner symplectic structure of ECH gravity encodes information about the corner coframe and metric, and revealed, following \cite{Freidel:2015gpa,Freidel:2019ees },  the non-commutativity and $\sll(2,\mathbb{R})_\para$ structure of the latter.

One should recall that, in the first place, the very existence of the corner symmetry charges  is due to the breaking of gauge symmetry at the boundary of the canonical slice. For instance, we have seen in \cite{Freidel:2020xyx} that canonical GR has vanishing surface boost charges at the corner, meaning that it is trivially boost-invariant at the corner. On the other hand, in the EH formulation the presence of the corner breaks boost-invariance, as revealed by the presence of non-vanishing corner charges for $\sll(2,\mathbb{R})_\perp$. Similarly, EH and canonical GR possess trivial charge associated with Lorentz invariance, while in the ECH tetrad formulation the corner breaks local Lorentz invariance and gives rise to non-vanishing charges.
In other words different formulation of the same theories have different notion of corner symmetries. 

\subsection{Conceptual motivations}

While it would seem that everybody agrees on these facts, there is a conceptual confusion on how to interpret them and deal with them. After all (continuing on the example of Lorentz transformations) one could proceed as in \cite{DePaoli:2018erh,Oliveri:2019gvm} and introduce the relative $\Theta_{\DPS}$ potential in order to cancel the Lorentz charges and thereby restore gauge invariance. But this restoration is then at the price of loosing access to the underlying charges, which is unfortunate since they give us an handle on quantization, and are the building blocks of the LQG quantization. More generally, we are arguing for a quantization program of gravity which uses the various charges of corner symmetry as the fundamental building blocks, so we should give them a more prominent role instead of taming them.
As argued in \cite{Freidel:2020xyx}, in this quantization program based on local holography, one should look for the maximally extended theory which reveals all the corner charges as element of the quantum geometry.  In this section we clarify these points and emphasize that we can \textit{both} restore gauge invariance and also have non-trivial corner charges. They key is to allow for the introduction of edge modes.

The historical view on this question \cite{Regge:1974zd,Balachandran:1993tm,Balachandran:1994up,Balachandran:1995qa,Carlip:1994gy,Carlip:2005zn} has been to simply accept that the presence of boundaries breaks gauge invariance. Recall that this breaking of gauge invariance is manifest once we realize that the canonical generator of gauge transformations does not vanish for transformations that are non-vanishing at the slice's boundary. For instance, in ECH gravity the generator of Lorentz gauge transformations associated with a gauge parameter $\alpha$ is given by
\be
\CH_\text{ECH}[\alpha]=\f{1}{2}\int_\Sigma E_{IJ}\wedge\rd_\omega\alpha^{IJ}.
\ee
These charges do not vanish after imposition of the Gauss law, but instead satisfy
\be\la{GaussLaw}
\CH_\text{ECH}[\alpha]\simeq\CH_\text{ECH}[\alpha'],
\q
\mathrm{when}
\q\alpha\stackrel{S}{=}\alpha',
\ee
where $\simeq$ means that the Gauss law is imposed. This means that generators can be separated in \textit{gauge} generators, associated with parameters $\alpha\stackrel{S}{=}0$ that vanish at the corner, and \textit{symmetry} generators, associated with transformations that do not vanish on $S$. The Gauss law \eqref{GaussLaw} is the expression of the fact that these canonical charges are independent of their bulk extension. The algebra generated by these charges is denoted by $\mathfrak{g}_S$, and the corresponding group by $G_S$. In the case of ECH gravity we have seen that $G_S=\mathrm{Diff}(S)\ltimes\big(\SL(2,\C)^S\times\SL(2,\mathbb{R})_\para^S\big)$. The presence of non zero-charges attached to the corner suggests that there are new corner degrees of freedom associated with the presence of the broken gauge group \cite{Carlip:1995cd,Strominger:2017zoo}. These degrees of freedom should play the role of boundary Goldstone modes for the broken symmetry. However, these Goldstone modes are not included in the usual treatments of gauge theories at boundaries, as they are implicitly superselected. As we are about to argue, we need a framework where they are explicitly present.

A more modern take on this, introduced in \cite{Donnelly:2016auv} and developed in \cite{Speranza:2017gxd,Geiller:2017xad,Geiller:2017whh, Freidel:2018fsk, Setare:2018mii, Geiller:2019bti}, consists in explicitly introducing new corner degrees of freedom, called \textit{edge modes}, in order to restore gauge invariance and decouple the notion of corner symmetry from that of gauge. The advantage is that we can, in this case, construct two sets of canonical charges, which for the example of Lorentz transformations with parameter $\alpha$ we denote $\CC[\alpha]$ and $\CQ[\alpha]$. The charge $\CC[\alpha]$ is the generator of gauge transformations. It vanishes on-shell and it generates a transformation of both the bulk and corner variables. The bulk component of $\CC[\alpha]$ imposes the Gauss law, while its corner component imposes a continuity equation relating the bulk and corner field. The charge $\CQ[\alpha]$, on the other hand, is the corner symmetry generator which acts on the corner variables only. It does not vanish on-shell, and is gauge-invariant in the sense that it commutes with the gauge generator $\CC[\alpha]$. One can then go back to the ``usual'' picture mentioned above if one explicitly breaks the corner gauge invariance. The residual symmetry which preserves this explicit breaking is a combination of the gauge transformation and the corner symmetry, and its generator $\CH[\alpha]$ is therefore given by a linear combination of $\CC[\alpha]$ and $\CQ[\alpha]$.

\subsection{Concrete implementation}

Let us now explain how to implement this construction with the explicit example of BF theory and then ECH gravity. For BF theory, we have in the bulk of a slice $\Sigma$ the canonical pair $(\tB^{IJ},\tilde{\omega}^{IJ})$.  In order to resolve once and for all the 
confusion between having a $B$-field which is commutative when viewed as a bulk variable, and non-commutative when viewed as a corner variable, 
we propose to introduce edge modes variables.
This means that we are considering corner variables denoted $(\BB^I,\e^I,\n^I,\vphi)$, where the straight bold font is used to emphasize that these corner edge mode fields are a priori independent from the bulk fields. These fields satisfy the constraints\footnote{These are just kinematical constraints which ensure that we have a 12-dimensional phase space, and their presence has nothing to do \textit{per se} with the fact that we are talking about edge modes.}
\be
\BB^I\n_I=0,
\q
\e^I\n_I=0,
\q
\n^I\n_I=\sigma.
\ee
Here $\BB^I$ is a vector-valued 2-form on $S$, $\n^I$ is an internal vector, $\e^I$ is a vector-valued 1-form on $S$, and $\vphi\in\SL(2,\mathbb{C})^S$ is an element of the corner symmetry group. Since these variables live at the corner, we drop the tilde label on them. The corner variables $(\BB^I,\e^I,\n^I)$ have a different status than $\vphi$. They are gauge-invariant, while $\vphi$ allows us to specify the choice of the corner \textit{gauge frame}. With these new corner edge mode fields we can now introduce the notion of \textit{extended symplectic potential} \cite{Donnelly:2016auv, Freidel:2019ofr,Geiller:2017xad,Geiller:2019bti}. 

The extended symplectic potential is obtained by subtracting from the original symplectic potential a corner potential for the edge modes. More precisely, in the example at hand it is
\be
\Theta_\BF^\ext\coloneqq\Theta_\BF-\bTh_\BF^S,
\ee
where $\Theta_\BF$ is given by \eqref{bare BF potential} or \eqref{BF decomposition}, and the corner symplectic potential for the edge modes is
\be\la{edge mode potential}
\bTh_\BF^S\coloneqq-\int_S\left(\sigma\BB_I\bd\n^I+\f{\beta}{2}\e_I\wedge\bd\e^I\right).
\ee
Comparing this corner potential with the corner potential $\Theta_\BF^S$ \eqref{BF corner potential} contained in $\Theta_\BF$ reveals two differences (aside from the innocent fact that we have chosen convenient notations for the BF coframes). First, the fields in $\bTh_\BF^S$ are edge-modes:corner variables with no bulk potential. Second, the edge mode corner potential is defined with a variational derivative $\bd$. This is an \textit{horizontal derivative} which depends on the corner edge mode field $\vphi$. It is given by
\be
\bd\coloneqq\delta-\delta_\chi,
\q
\mathrm{with}
\q
\chi\coloneqq\vphi^{-1}\delta\vphi.
\ee
Here $\delta_\alpha$ represents as usual the gauge transformation associated with a gauge parameter $\alpha$. For this notion of horizontal derivative the gauge parameter is the Lie algebra-valued variational 1-form $\chi=\vphi^{-1}\delta\vphi$, which can be interpreted as a flat connection on field space. The explicit action of $\bd$ on $\e^I$ for instance is simply $\bd\e^I=\delta\e^I+\chi^I{}_J\e^J$. The covariant variational derivative\footnote{We could also extend the horizontal field derivative to be covariant under the Diff$(S)$ symmetry group by introducing a position frame as in \cite{Donnelly:2016auv,Speranza:2017gxd}. In this case  everything said in this section for $\SL(2,\C)^S$ can be extended to $\mathrm{Diff}(S)\ltimes\SL(2,\C)^S$. For this paper we emphasize only the $\SL(2,\C)^S$ modes which are key for commutativity. } was first introduced in \cite{Donnelly:2016auv}, and conceptualized as an horizontal field derivative in \cite{Gomes:2016mwl}. Like $\delta$, the covariant variation $\bd$ satisfies the Cartan axiom $\bd^2=0$.

A conceptually very important point for what follows is to acknowledge that the corner fields $(\BB^I,\e^I,\n^I)$ are a priori different from the corresponding bulk fields. They become only identified once gluing conditions are imposed. This is key to the mechanism of restoration of gauge invariance. To see this, let us consider the bulk gauge transformation $\delta_\alpha\omega^{IJ}=\rd_\omega\alpha^{IJ}$ and $\delta_\alpha B^{IJ}=[B,\alpha]^{IJ}$. We choose the edge mode fields to be such that the extension of the gauge transformation at the corner is
\be
\delta_\alpha\n^I=\delta_\alpha\e^I=\delta_\alpha\BB^I=0,
\q
\delta_\alpha\vphi=-\alpha\vphi.
\ee
This means that the corner fields $(\BB^I,\e^I,\n^I)$ are gauge invariant observables. The generator $\CC[\alpha]$ mentioned above is the canonical generator of these extended gauge transformations. It is defined as usual with the covariant phase space formula
\be
\delta\CC_\BF[\alpha]=-\delta_\alpha\ip\Omega_\BF^\ext,
\ee
from which we find
\be\la{extended gauge generator}
\CC_\BF[\alpha]=\CH_\BF[\alpha]-\f{1}{2}\int_S(\vphi^{-1}\alpha\vphi)^{IJ}\JJ_{IJ}=\f{1}{2}\int_\Sigma B_{IJ}\wedge\rd_\omega\alpha^{IJ}-\f{1}{2}\int_S(\vphi^{-1}\alpha\vphi)^{IJ}\JJ_{IJ}.
\ee
Here we have introduced the corner generator 
\be\la{cJJ}
\JJ_{IJ}\coloneqq2\sigma\BB^{[I}\n^{J]}+\beta(\e\wedge\e)^{IJ}.
\ee
The corner symplectic potential implies that these generators satisfy an ultra-local Lorentz algebra
\be 
\{\JJ_{IJ}(x),\JJ_{KL}(y)\}=\big(\eta_{JK}\JJ_{IL}+\eta_{IL}\JJ_{JK}-\eta_{IK}\JJ_{JL}-\eta_{JL}\JJ_{IK}\big)(x)\delta^2(x,y).
\ee
Now, demanding the validity of the extended Gauss law $\CC_\BF[\alpha]\simeq0$ imposes the bulk and corner constraints
\be\la{gluing}
\rd_\omega B^{IJ}\stackrel{\Sigma}{\simeq}0,
\q
B^{IJ}\stackrel{S}\simeq(\vphi\JJ\vphi^{-1})^{IJ}.
\ee
The corner constraint is a boundary condition relating the pull-back of the bulk fields on $S$ with the value of the corner fields. Importantly, this identification involves an element of the corner symmetry group $\vphi\in\SL(2,\mathbb{C})^S$. The reason  this group element is needed is that
the naive identification $B^{IJ}\stackrel{S}{\simeq}\JJ^{IJ}$ is second class while the identification $B^{IJ}\stackrel{S}\simeq(\vphi\JJ\vphi^{-1})^{IJ} $ is first class. A proof of this is given in Appendix \ref{B gluing}. The first class nature of the constraint ensures that we can let go of the notion of boundary $B$ field, and replace it by the dressed Lorentz generator. Since $\JJ^{IJ}$ is naturally a non-commutative corner variable, this resolve the confusion about the difference of Poisson structure of the bulk and corner $B$ fields. This is how the puzzle pointed out in Section \ref{sec:danger} is resolved by the introduction of edge modes.

Imposing the bulk and corner Gauss constraints \eqref{gluing} means that $\CC_\BF[\alpha]\simeq0$, and therefore that we have restored gauge invariance. Now that gauge invariance has been restored by the addition of the corner degrees of freedom, we can construct a gauge-invariant, and therefore physical charge. This charge is defined as being associated with transformations $\Delta_\alpha$ which leave the bulk fields unchanged, i.e. $\Delta_\alpha(B^{IJ},\omega^{IJ})=0$, while rotating the corner fields as
\be\la{rotb}
\Delta_\alpha\BB^I=-\alpha^I{}_J\BB^J,
\q
\Delta_\alpha\e^I=-\alpha^I{}_J\e^J,
\q
\Delta_\alpha\n^I=-\alpha^I{}_J\n^J,
\q
\Delta_\alpha\vphi=0.
\ee
The canonical generator for this transformation is the object $\CQ[\alpha]$ mentioned above. Here it is the corner charge associated with $\JJ$ introduced in \eqref{cJJ}, i.e.
\be
\CQ_\BF[\alpha]=\Delta_\alpha\ip\bTh^S_\BF=\f{1}{2}\int_S\alpha^{IJ}\JJ_{IJ}.
\ee
Because of the definition of its action, this charge comes entirely from the edge mode symplectic structure. Furthermore, it is gauge-invariant in the sense that $\{\CC_\BF[\alpha],\CQ_\BF[\beta]\}=0$, and it enters the relationship between the extended gauge generator $\CC_\BF[\alpha]$ and the ``naive'' charge $\CH_\BF[\alpha]$ given by
\be
\CH_\BF[\alpha]=\CC_\BF[\alpha]+\CQ_\BF[\vphi^{-1}\alpha\vphi],
\ee
as can be seen from \eqref{extended gauge generator}. This shows that the naive transformation generated by $\CH_\BF[\alpha]$ is in fact a gauge transformation followed by a corner rotation, which explains why its generator does not vanish. Note that the rotation \eqref{rotb} of the corner generators which leaves $\vphi$ fixed is equivalent to a right translation of the \textit{gauge frame} $\vphi$. This follows from the fact that $(\Delta_\alpha+\tilde{\Delta}_\alpha)\ip\bd=0$ for the right translation defined\footnote{This is the notion of surface symmetry which was actually initially used and named $\Delta_\alpha$ in \cite{Donnelly:2016auv,Speranza:2017gxd,Geiller:2017xad,Geiller:2017whh, Freidel:2018fsk}.} as $\tilde{\Delta}_\alpha(\BB^I,\e^I,\n^I)=0$ and $\tilde{\Delta}_\alpha\vphi=\vphi\alpha$. Equivalently, we can say that the corner transformations
\be
(\Delta_\alpha+\tilde{\Delta}_\alpha)\big(\BB^I,\e^I,\n^I,\vphi\big)=\big(\BB^J\alpha_J{}^I,\e^J\alpha_J{}^I,\n^J\alpha_J{}^I,\vphi\alpha\big)
\ee
are pure gauge \textit{and} the corresponding canonical generator identically vanishes.

As we have discussed previously at length, the case of ECH gravity is obtained from the BF analysis by imposing the simplicity constraints. In the bulk they are $B^{IJ}=E^{IJ}[e]$, and the \textit{boundary simplicity constraint} is
\be
\BB^I=\f{1}{2}(\e\times\e)^I.
\ee
It is important to appreciate that the knowledge of $\BB^I$ only determines $\e^I$ up to an $\SL(2,\R)^S$ rotation $\e_a^I\to\rho_a{}^b\e_b^I$ where $\det\rho=1$. This means that, after imposing the bulk and corner simplicity constraints, the gluing condition \eqref{gluing} can be written in terms of the coframe field as
\be\la{frame gluing}
\te_a^I\stackrel{S}{\simeq}\rho_a{}^b\e_b^J\varphi_J{}^I=(\e\cdot\bvphi)_a^I,
\ee
where we have introduced the corner group element $\bvphi\coloneqq(\rho,\vphi)\in G_S$, with here the corner symmetry group $G_S=\SL(2,\R)^S\times \SL(2,\C)^S$.

To understand the physical meaning of identity \eqref{frame gluing}, let us consider two closed regions $\Sigma_L$ and $\Sigma_R$ sharing a common corner $\Sigma_L\cap\Sigma_R=S_{LR}=\pa\Sigma_L\cap\pa\Sigma_R$, and further assume that the bulk field is continuous across the corner. This continuity equation implies a matching condition of the corner fields which involves an holonomy attached to each point of the sphere. Denoting by $(\te_L,\te_R)$ the bulk coframe fields associated with the left and right regions, and by $(\e_L,\e_R)$ the corner ones, the continuity equation takes the form
\be
\te_L\stackrel{S_{LR}}{=}\te_R
\q\iff
\q\e_L\stackrel{S_{LR}}=\e_R\cdot\bvphi_{LR},
\ee
where $\bvphi_{LR}\in G_S$ is the corner holonomy with value in the extended corner symmetry group. The $\SL(2,\C)^S$ component of this holonomy can be understood as a change of section of the coframe bundle when we go from the left to the right region. The $\SL(2,\R)^S$ component, on the other hand, is a fundamentally new contribution whose important interpretation will be developed in a forthcoming publication.

\subsection{Trivial limits}

From the general perspective presented here, it is now easy to recover the naive picture where only the charge $\CH_\BF[\alpha]$ appears. This is done by simply choosing the corner group element to be the identity. The condition $\vphi=1$ effectively breaks the gauge invariance at the corner. The transformation generated by $\CH_\BF[\alpha]=\CC_\BF[\alpha]+\CQ_\BF[\alpha]$ is then the only one which preserves the condition $\vphi=1$. In this gauge, the gluing condition becomes $\te^I\stackrel{S}{\simeq}\e^I$ and there is no need to introduce independent corner fields different from the bulk ones.

We can also recover the approach of \cite{DePaoli:2018erh,Oliveri:2019gvm}. What is done in these references can be interpreted in our framework as imposing strongly, independently of any corner Gauss law, the condition $\te^I\stackrel{S}{=}\e^I$ in the extended symplectic potential. Imposing this condition strongly also forces $\vphi=1$ and implies, by virtue of \eqref{EH/GR1} and \eqref{EHsympdec}, that the extended potential reduces to the gravity potential
$\Theta_\ECH^\ext=\Theta_\GR-\delta\left(\int_\Sigma\teps\tK\right)$. This choice then amounts to working with the metric gravitational potential which has vanishing Lorentz charges. This way of restoring gauge invariance kills all the non-trivial charges and it corresponds at the quantum level to a choice of representation which reduces the effective symmetry group. As mentioned above however, the reason for which we cannot impose strongly (and innocently) the condition $\te^I\stackrel{S}{=}\e^I$ comes from the fact that $\te^I$ is a commuting field in the bulk while $\e^I$ is a non-commuting corner field. The condition $\te^I\stackrel{S}{=}\e^I$ is therefore a second class constraint. The boundary condition $B^{IJ}\stackrel{S}{\simeq}(\vphi\JJ\vphi^{-1})^{IJ}$ or \eqref{frame gluing} is, on the other hand, a first class constraint due to the presence of the corner gauge frame $\vphi$ and the Gauss law. This is one of the main points made in \cite{Donnelly:2016auv}, where it served as the motivation for introducing the edge modes.

Some authors \cite{Gomes:2018shn,Gomes:2018dxs} have also pointed out that it is possible to restore gauge invariance without introducing new corner fields, and have concluded from this fact that edge modes do not exists or are not relevant. Once again such perspective easily follows from the general framework presented here. One can effectively kill all the corner charges while keeping gauge symmetry by imposing some gauge invariant boundary condition on $\vphi$, such as $\vphi=\vphi(A)$, where $\vphi(A^g)=\vphi(A) g$ is a functional of $A$ which intertwines the gauge transformation of the potential with the right translation of the gauge frame (see \cite{Vilkovisky,Zwanziger,Lavelle:1995ty} for explicit constructions). This type of conditions effectively break the corner symmetry and the only charges that commute with the condition $\vphi=\vphi(A)$ are the gauge charges $\CC[\alpha]$. Of course, the fact that it is possible to fix the corner symmetry doesn't mean that the corner edge modes do not enter the counting of degrees of freedom, it only means that the corner edge modes can be rotated into one another by a corner symmetry.

While it might be harmless to ignore edge modes at the  classical level, it is not at the quantum level. Edge modes reveal all their usefulness and efficacy, as we demonstrate in this series of works, once we take into account the presence of non-trivial commutation relations. The main point is that, although bulk fields and edge mode fields look the same, they have \textit{different} commutation relations and \textit{cannot} be identified as operators.

We take the point of view that a symmetry algebra needs to be represented non trivially in order to access all degrees of freedom. After all, the central insight of the LQG description of quantum geometry is to keep track precisely of the flux charges labelled by $\SU(2)$ spins coloring spin networks, which are in fact coming from the rotational edge modes and satisfy a non-trivial commutation relation. Here we push this logic further, and reveal that it inevitably leads to additional charges and new quantum numbers on the corner.

\section{Conclusion}
\la{conclusions}

Our proposal for quantum gravity is rooted in the notion of representation of the corner symmetry algebra. As a step towards this construction, we have here analyzed tetrad gravity by shifting the emphasis from the bulk to the corner: Instead of working with a connection in the bulk, and having to deal with the ambiguities associated to it, here we have instead considered the Holst term on the corner and kept the bulk parametrized by the universal GR symplectic structure. This is still a non-trivial construction because the Holst term and the associated coupling, which is the Barbero--Immirzi parameter, now control the non-commutativity of the corner frame field and, as we showed in Section \ref{sec:bmetric}, the non-commutativity of the corner metric as well. Shifting the emphasis to the corner also enables us to resolve  the tension of having a commutative bulk $B$ field while $B$ becomes non-commutative at the corner (after imposing Gauss Law).

To summarize, in this paper we have achieved three results. The first set of results is technical and follows from a careful decomposition of $B$ field and the connection in terms of tangential/normal and horizontal/vertical components. This led us to the introduction of the notion of BF coframes and to a new split of the BF symplectic potential in a bulk and corner piece. We have also analyzed in details the simplicity constraints in this context, and presented the construction of the relative corner potential relating the Einstein--Cartan--Holst formulation to the ADM and Einstein--Hilbert metric formulations of gravity. These results reorganize in a coherent manner a large collection of disparate results in the literature on first order gravity and its symplectic potential \cite{Peld_n_1994, Thiemann:2001yy, Mondragon:2004nw,Liu:2009em, Engle:2010kt, Engle:2011vf, Bodendorfer:2013sja, Corichi:2013zza,  Bodendorfer_2013,  Bodendorfer:2013jba,   DePaoli:2018erh, Oliveri:2019gvm}.

The second result consists in showing, starting from a new parametrization of the corner phase space given in \eqref{TS2}, that the corner metric satisfies an ultra-local $\sll(2,\R)^S$ algebra. This generalizes the conclusions of \cite{Freidel:2015gpa, Freidel:2016bxd, Freidel:2019ees} in a context where the time gauge is not imposed and therefore full Lorentz covariance preserved. The quantization of this algebra led us to the conclusion that the area spectrum is quantized even in the presence of full Lorentz symmetry. This result resolves a long-standing debate in the quantum gravity community, which revolved around the possibility to reconcile Lorentz invariance with the discreteness of the area spectrum \cite{Alexandrov:2000jw, Alexandrov:2002br, Alexandrov:2001pa, Alexandrov:2001wt, Livine:2002ak, Freidel:2002hx, Rovelli:2002vp, Dittrich:2007th, Rovelli:2007ep, Alexandrov:2010un, Rovelli:2010ed, Achour:2013gga}. Further insights into this reconciliation are provided in \cite{Edge-Mode-III}. A similar conclusion was reached by Wieland using a corner algebra associated with null boundaries and written in spinor variables \cite{Wieland:2017cmf}. It would be interesting to understand the relationship between the two frameworks. Since our work exclusively used space-like or time-like slices, this suggests that there should be a more general framework generalizing our results to all types of slices, including the null ones.

The third result is more conceptual. We have explained how to resolve the fundamental tension which exists between bulk and corner variables. Classically, and if we ignore the canonical structure, one can think of the corner variable as being obtained by continuity as the pull-back of the bulk field on the corner. But this is not possible at the quantum level or even at the classical level if we take into account the canonical structure. The issue is that the pull-back of the bulk fields possess \textit{different} commutation relations than the corner fields. We have explained how this tension is resolved by the introduction of edge modes carrying an independent corner symplectic potential. To do so, we have replaced the condition of continuity by the imposition of gauge invariance extended to the corner. The tension between the different commutation relation is resolved by the presence of corner gauge frames that relate bulk and boundary fields. This result conceptualizes and extends the work done earlier in \cite{Donnelly:2016auv,Freidel:2019ofr}.

Although the analysis presented in this paper is semi-classical, it prepares the terrain for a new form of quantization of the gravitational variables focusing on their expression as corner charges of symmetry. 

\section*{Acknowledgement}

We would like to thank Simone Speziale for discussions. Research at Perimeter Institute for Theoretical Physics is supported in part by the Government of Canada through NSERC and by the Province of Ontario through MRI.

\appendix

\section{Notations and conventions}
\la{appendix1}

Throughout this work we consider that the spacetime $M$ is equipped with a Lorentzian metric $g_{\mu\nu}$ of signature $(-,+,+,+)$. We use $\mu,\nu,\dots$ to denote spacetime indices, while $a,b,\dots$ when this are pulled back on the co-dimension 2 corner $S$. The internal metric is $\eta_{IJ}=\text{diag}(-1,1,1,1)$, and we use $I,J,\dots$ to denote internal Lorentz indices. The these internal indices the Hodge duality operation is
\be
(*M)^{IJ}=\f{1}{2}{\eps^{IJ}}_{KL}M^{KL},\q*^2=-1.
\ee
We denote the antisymmetrisation of indices with the bracket $[A_1\cdots A_n]$, and the convention that $[[A_1\cdots A_n]]=[A_1\cdots A_n]$. In particular, in the case of two indices we have
\be\la{antisymmetrisation}
M_{[IJ]}=\f{1}{2}(M_{IJ}-M_{JI}),\q M_{[IJ]}N^{[IJ]}=M_{[IJ]}N^{IJ}.
\ee
The commutator of Lie algebra-valued forms is denoted by
\be\la{def commutator}
[M\wedge N]^{IJ}={M^I}_K\wedge N^{KJ}-{M^J}_K\wedge N^{KI}=2{M^{[I}}_K\wedge N^{KJ]}.
\ee
This commutator is such that 
\be
 *[M\wedge N]= [*M\wedge N]= [M\wedge *N].
\ee
We have the relations
\begin{subequations}
\be
\eps_{IJKL}\eps^{MNPQ}&=-24\delta^{[M}_I\delta^N_J\delta^P_K\delta^{Q]}_L,\\
\eps_{IJKL}\eps^{INPQ}&=-6\delta^{[N}_J\delta^P_K\delta^{Q]}_L=-2(\delta^{N}_J\delta^{[P}_K\delta^{Q]}_L+\delta^{Q}_J\delta^{[N}_K\delta^{P]}_L+\delta^{P}_J\delta^{[Q}_K\delta^{N]}_L),\\
\eps_{IJKL}\eps^{IJPQ}&=-4\delta^{[P}_K\delta^{Q]}_L=-2(\delta^P_K\delta^Q_L-\delta^Q_K\delta^P_L),\\
\eps_{IJKL}\eps^{IJKQ}&=-6\delta^Q_L.
\ee
\end{subequations}
With the internal normal $n^I$ such that $n^In_I=\sigma$ we define 
\be 
\teps_{IJK}\coloneqq\eps_{IJKL}n^L.
\ee
These satisfy
\begin{subequations}
\be
\teps_{IJK}\teps^{ILM}&=-\sigma(\tilde{\delta}^L_J\tilde{\delta}^M_K-\tilde{\delta}^M_J\tilde{\delta}^L_K),\\
\teps_{IJK}\teps^{IJM}&=-2\sigma\tilde{\delta}^M_K,
\ee
\end{subequations}
where $\tilde{\delta}^M_K\coloneqq\delta^M_K-\sigma n_Kn^M$. Finally, we define the cross product
\be
(M\times N)_I\coloneqq\teps_{IJK}M^J\wedge N^K,
\ee
between vector-valued forms.
The cross product satisfy
\be
-\f{\sigma}2\teps^{IJ}{}_K(M\times M)^K=M^I\wedge M^J=(M\wedge M)^{IJ}.
\ee

\section{Gravitational equations of motion}
\la{gravityeoms}

\subsection{Torsion equations}

Here we want to decompose the torsion equations, and explain the geometrical meaning of the decomposition \eqref{decomposition of omega} of the connection. More precisely, we are going to show that if $\omega^{IJ}$ is the spin connection preserving $e^I$, then the pull-back of $\Gamma$ to $\Sigma$ is the spin connection preserving the pull-back $\te^I=e^I-\sigma \un n^I$. In order to see this, we first project the torsion $T^I=\rd_\omega e^I$ and use $\rd_\omega n^I=K^I$ and $\rd_\Gamma n^I=0$ to obtain
\be
n_IT^I=\rd\un+e_I\wedge K^I\stackrel{\Sigma}{=}e_I\wedge K^I\simeq0,
\ee
where the weak equality $\simeq$ means that we have imposed the vanishing of the torsion $T^I\simeq0$. By virtue of \eqref{covariant Gamma of n} we have
\be
\rd_\Gamma e^I=\rd_\Gamma(\te^I+\sigma\un n^I)=\rd_\Gamma\te^I+\sigma\rd\un n^I-\sigma\un\wedge\rd_\Gamma n^I=\rd_\Gamma\te^I+\sigma\rd\un n^I.
\ee
Similarly to the decomposition of forms \eqref{horizontal/vertical}, we can also decompose the differential of an horizontal form as
\be
\rd_\Gamma\tilde{\alpha}=\tilde{\rd}_\Gamma\tilde{\alpha}+\sigma\un\wedge\CL^{\va\Gamma}_{\hat{n}}\tilde{\alpha},
\ee
where $\tilde{\rd}$ is the pull-back differential on $\Sigma$ and we have introduced the covariant Lie derivative
\be
\CL^{\va\Gamma}_\xi\coloneqq\xi\ip(\rd_\Gamma\,\cdot)+\rd_\Gamma(\xi\ip\cdot).
\ee
This derivative is such that $\CL^{\va\Gamma}_\xi n^I=0$. We can also decompose the differential of $\un$ using
\be
\rd\un=\tilde{\rd}\un+\sigma\un\wedge\CL_{\hat{n}}\un.
\ee
From now on we will also use the hypersurface orthogonality condition $\tilde{\rd}\un=0$, which follows from the fact that $\un$ is normal to the slices $\Sigma$.

We can use these ingredients to decompose the torsion into horizontal and vertical components. Starting from the decomposition of $\omega$ \eqref{decomposition of omega}, we have
\be\la{torsion decomposition}
T^I
&=\rd_\Gamma e^I-\sigma\un\wedge K^I+\sigma n^I(e_J\wedge K^J)\cr
&=\rd_\Gamma\te^I-\sigma\un\wedge K^I+\sigma n^I(\rd\un+\te_J\wedge K^J)\cr
&=\tilde{\rd}_\Gamma\te^I-\sigma\un\wedge\big(K^I-\CL^{\va\Gamma}_{\hat{n}}\te^I\big)+\sigma n^I(\rd\un+\te_J\wedge K^J)\cr
&=\tilde{\rd}_\Gamma\te^I-\sigma\un\wedge\big(\tK^I-\CL^{\va\Gamma}_{\hat{n}}\te^I\big)+\sigma n^I\te_J\wedge\tK^J+n^I\un\wedge\big(\CL_{\hat{n}}\un-\te_J\wedge K_n^J\big),
\ee
where we have used $K^I=\tK^I+\sigma\un\wedge K_n^I$, with $K_n^I\coloneqq\hat{n}\ip K^I$ the acceleration vector, and we have used the hypersurface orthogonality condition $\tilde{\rd}\un=0$. This enables us to read the horizontal and vertical components of the torsion equations. The horizontal equations are
\be\la{horizontal torsion}
\tilde{\rd}_\Gamma \te^I\simeq0,
\q
\te_I\wedge \tK^I\simeq0.
\ee
The first condition tells us that the pull-back of $\Gamma$ is the spin connection of $\te$. The second condition means that the tensor $\tK^{IJ}$ entering the expansion $\tK^I=\tK^{IJ}\te_J$ is symmetric. On the other hand, the vertical equations are
\be\la{vertical torsion}
\CL^{\va\Gamma}_{\hat{n}}\te^I\simeq\tK^I,
\q
\te_I\wedge K_n^I\simeq\CL_{\hat{n}}\un.
\ee
The first one tells us that the pull-back of $K^I$ can be understood as the extrinsic curvature, i.e. the normal derivative of the induced coframe.

\subsection{Einstein equations}
\la{sec:EE equations}

Here we decompose the first set of equations of motion in \eqref{EC EOMs}, which are Einstein's equations in tetrad variables. Let us start by using the normal/tangential decomposition of the Lorentz curvature to write the Einstein tensor as
\be
G^I
&=(*+\beta)\left(F^{IJ}_\para+2\sigma F^{[I}_\per n^{J]}\right)\wedge e_J\cr
&=(*F_\para)^{IJ}\wedge e_J+\beta F_\para^{IJ}\wedge e_J-\sigma(F_\per\times e)^I+\sigma\beta F^I_\per\wedge\un-\sigma\beta n^IF^J_\per\wedge e_J.
\ee
We can then insert the normal and tangential components \eqref{components of Lorentz F} of the Lorentz curvature tensor in this expression. They are given by
\be\la{normal/tangent F}
F_\per^I=\rd_\Gamma K^I,
\q
F_\para^{IJ}=R^{IJ}(\Gamma)-\sigma(K\wedge K)^{IJ}.
\ee
Using these components, contracting the Einstein tensor with the internal normal, and denoting $R^I(\Gamma)=\teps^{IJK}R_{JK}(\Gamma)$, we get the normal equations
\be
G_\per\coloneqq G^In_I=-R^I(\Gamma) \wedge e_I+\f{\sigma}{2}(K\times K)^I\wedge e_I-\beta\rd_\Gamma K^I\wedge e_I\approx0.
\ee
The horizontal component of this equation gives the \textit{Hamiltonian constraint}
\be
C\coloneqq-\tilde{R}^I(\tilde\Gamma)\wedge\te_I+\f{\sigma}{2}(\tK\times\tK)^I\wedge\te_I-\beta\tilde{\rd}_\Gamma\tK^I\wedge\te_I.
\ee
The remaining components of the Einstein tensor (i.e. the ones which have been killed by the normal projection above) are given by
\be
G^I_\para=\beta R^{IJ}\wedge e_J-\sigma\beta(K\wedge K)^{IJ}\wedge e_J-\sigma(\rd_\Gamma K\times e)^I+\sigma\beta\rd_\Gamma K^I\wedge\un\approx0.
\ee
This equation can be considerably simplified by taking the pull-back to $\Sigma$ and using the relations obtained in \eqref{horizontal torsion}, namely $\tilde{\rd}_\Gamma\te^I\simeq0$ and $\te_I\wedge\tK^I\simeq0$. It yields the conservation law
\be
\tilde{\rd}_\Gamma \tP^I\approx0,
\ee
where $\tP^I\coloneqq-\sigma(\tK\times\te)^I$ is the momentum aspect.

\section{Various proofs}
\la{appendix3}

\subsection{Gauge transformations of the decomposed connection}
\la{appendix:gauge}

In this appendix, we give for completeness the behavior of the different parts of the Lorentz connection under Lorentz gauge transformations. In particular, this shows that $\Gamma^{IJ}$ transforms as a connection and $K^I$ as a tensor.

Using the decomposition \eqref{tensor decomposition}, we can write the Lorentz gauge parameter as
\be
\alpha^{IJ}=\bar{\alpha}_\per^{IJ}+\bar{\alpha}_\para^{IJ},
\q
\bar{\alpha}_\per^{IJ}\coloneqq 2\sigma\alpha_\per^{[I}n^{J]},
\q
\bar{\alpha}_\para^{IJ}\coloneqq-\sigma\teps^{IJ}{}_{K}\alpha_\para^K .
\ee
Here we have used a notation where $\bar \alpha_\per$ denotes a Lie algebra element and $\alpha_\per$ a vector perpendicular to $n_I$. By definition, $\alpha_\per^I$ represent the change in the normal vector since $\delta_\alpha n^I=-\alpha^{IJ}n_J=-\alpha_\per^I$. We can use this to decompose the gauge transformation as
\be
\delta_\alpha\omega^{IJ}=\rd_\omega\alpha^{IJ}=\delta_{\bar{\alpha}_\per}\omega^{IJ}+\delta_{\bar{\alpha}_\para}\omega^{IJ}.
\ee
Using the decomposition \eqref{decomposition of omega}, we can then write the two gauge transformation on the right-hand side as
\be
\delta_{\bar{\alpha}_\per}\omega^{IJ}
&=\rd_\omega\bar{\alpha}_\per^{IJ}\cr
&=\rd_\Gamma\bar{\alpha}_\per^{IJ}+\sigma(K^In_A-n^IK_A)\bar{\alpha}_\per^{AJ}+\sigma(K^Jn_A-n^JK_A)\bar{\alpha}_\per^{IA}\cr
&=\rd_\Gamma\bar{\alpha}_\per^{IJ}-2\sigma K^{[I}\alpha_\per^{J]},
\ee
and similarly for $\delta_{\bar{\alpha}_\para}\omega^{IJ}$. Using the fact that $\rd_\Gamma n^I=0$, one can rewrite these transformations as
\be\la{per para gauge omega}
\delta_{\bar \alpha_\per}\omega^{IJ}=2\sigma(\rd_\Gamma\alpha^{[I}_\per n^{J]}-  K^{[I}\alpha_\per^{J]}),
\q
\delta_{\bar{\alpha}_\para}\omega^{IJ}=2(K\times\alpha_\para)^{[I}n^{J]}-2\sigma*(\rd_\Gamma\alpha_\para^{[I}  n^{J]}).
\ee
From this we then get
\be\la{per para gauge omega n}
(\delta_{\bar{\alpha}_\per}\omega^{IJ})n_J=\rd_\Gamma\alpha^I_\per,
\q
(\delta_{\bar{\alpha}_\para}\omega^{IJ})n_J=\sigma(K\times\alpha_\para)^I.
\ee
On the other hand, using $\delta_\alpha n^I=-\alpha^{IJ}n_J=-\alpha_\per^I$,  the variation of the decomposition \eqref{decomposition of omega} leads to
\be\la{gauge transfo omega decomposed}
\delta_\alpha\omega^{IJ}=\delta_\alpha\Gamma^{IJ}+2\sigma\delta_\alpha K^{[I}n^{J]}-2\sigma K^{[I}\alpha_\per^{J]},
\ee
and using $(\delta_\alpha\Gamma^{IJ})n_J=\rd_\Gamma\alpha^I_\per$ gives
\be
(\delta_\alpha\omega^{IJ})n_J=\rd_\Gamma\alpha^I_\per+\delta_\alpha K^I-\sigma n^I\delta_\alpha K^Jn_J=\rd_\Gamma\alpha^I_\per+\delta_\alpha K^I-\sigma n^I\alpha^J_\per K_J.
\ee
Taking the normal and tangential components of $\alpha$ in this equation and comparing to \eqref{per para gauge omega n} then tells us that
\be
\delta_{\bar{\alpha}_\per}K^I=\sigma n^I\alpha^J_\per K_J,
\q
\delta_{\alpha_\para}K^I=\sigma(K\times\alpha_\para)^I.
\ee
This is in agreement with what we could have derived from the definition $K^I=\rd_\omega n^I$. Indeed, this gives
\be
\delta_\alpha K^I
&=\rd_\omega\delta_\alpha n^I+\delta_\alpha\omega^{IJ}n_J\cr
&=-\rd_\omega\alpha_\per^I+\rd_\omega\alpha^{IJ}n_J\cr
&=-\alpha^{IJ}\rd_\omega n_J\cr
&=-\alpha^{IJ}K_J\cr
&=\sigma n^I\alpha^J_\per K_J+\sigma(K\times\alpha_\para)^I.
\ee
Finally, comparing \eqref{per para gauge omega} with \eqref{gauge transfo omega decomposed} gives
\be
\delta_{\bar{\alpha}_\per}\Gamma^{IJ}=2\sigma\rd_\Gamma\alpha_\per^{[I}n^{J]},
\q
\delta_{\bar{\alpha}_\para}\Gamma^{IJ}=-2\sigma \teps^{IJ}{}_K \rd_\Gamma\alpha_\para^K,
\ee
which implies that
\be
(\delta_\alpha\Gamma)^I_\per=\rd_\Gamma\alpha_\per^I,
\q
(\delta_\alpha\Gamma)^I_\para=\rd_\Gamma\alpha_\para^I.
\ee
We therefore get as expected that $K$ transforms as a tensor and $\Gamma$ as a connection.

\subsection{Momentum aspect identities}
\la{appendix:PM}

Here we give two useful identities involving the momentum $\tP^I=-\sigma(\tK\times\te)^I$. First, we have that
\be
\tE_I\wedge\delta\tK^I
&=\f{1}{2}\teps_{IJK}(\te\wedge\te)^{JK}\wedge\delta\tK^I\cr
&=\f{1}{2}\te^J\wedge\delta(\teps_{IJK}\te^K\wedge\tK^I)-\f{1}{2}\teps_{IJK}\te^J\wedge\delta\te^K\wedge\tK^I-*(\te\wedge\te)_{IJ}\wedge\tK^I\delta n^J\cr
&=-\f\sigma2\te^I\wedge\delta\tP_I+\f{\sigma}{2}\delta\te^I\wedge\tP_I\cr
&=\sigma\tP_I\wedge\delta\te^I-\f{\sigma}{2}\delta(\te^I\wedge\tP_I),
\ee
where for the third equality we have used $\tK^In_I=0=n_I\delta n^I$ to write
\be\la{vanishing EK delta n}
*(\te\wedge\te)_{IJ}\wedge\tK^I\delta n^J=2\sigma\tE_{[I}n_{J]}\wedge\tK^I\delta n^J=0.
\ee
We now prove equation \eqref{momentum P identity}. Using the pull-backs of the tetrad and the curvature, we have
\be\la{metric P relation}
\int_\Sigma \tP_I\wedge M^I
&=-\sigma\int_\Sigma(\tK\times\te)_K\wedge M^K\cr
&=-\sigma\int_\Sigma\tK_\mu^I(\teps_{IJK}\teps^{\mu\nu\rho}\te_\nu^J)M_\rho^K\cr
&=\int_\Sigma(\det\te)\tK_\mu^I\big(\te_I^\mu\te_K^\rho-\te_I^\rho\te_K^\mu\big)M_\rho^K\cr
&=\int_\Sigma\teps\big(\tK\tilde{g}_\mu{}^\rho-\tK_\mu{}^\rho\big)M_\rho^K\te_K^\mu,
\ee
where we have used $\teps_{IJK}\teps^{\mu\nu\rho}\te_\nu^J=-\sigma(\det\te)(\te_I^\mu\te_K^\rho-\te_I^\rho\te_K^\mu)$ and $\te_\nu^I\te_I^\mu=\tilde{g}_\nu{}^\mu$.

\subsection{DPS variational 1-form identity}
\la{app:DPS}

Here we give the proof of \eqref{DPS identity}. In order to establish this we introduce  the internal and spacetime normal 2-forms
\be
\eps^\perp_{IJ}\coloneqq *(\tilde{\bar{e}}\wedge\tilde{\bar{e}})_{IJ}=\sqrt{q}(n_Is_J-s_In_J),\q\eps^\perp_{\mu\nu}\coloneqq e_\mu^Ie_\nu^J\eps^\perp_{IJ}=\sqrt{q}(n_\mu s_\nu-s_\mu n_\nu).
\ee
These 2-forms are such that
\be
\eps^\perp_{\mu\nu}n^\mu=-\sqrt{q}\,s_\nu,\q\eps^\perp_{\mu\nu}s^\mu=-\sqrt{q}\,n_\nu,\q\tbE_I=(\tilde{\bar{e}}\times\tilde{\bar{e}})_I=\eps^\perp_{IJ}n^J=\sqrt{q}\,s_I.
\ee
One can then show that 
\be
*( e\wedge e)_{IJ}\varpi^{IJ}
&=
*( e\wedge e)_{IJ}e^{I\mu}\delta e_\mu^J\cr
&\stackrel{S}{=}\eps^\perp_{IJ}e^{I\mu}\delta e_\mu^J\cr
&=\eps^\perp_{IJ}\big(s^I s^\mu - n^I n^\mu\big)\delta e_\mu^J\cr
&=\eps^\perp_{IJ}\big(s^I\delta s^J-n^I\delta n^J\big)-\eps^\perp_{IJ}\big(s^Ie^J_\mu\delta s^\mu-n^Ie^J_\mu\delta n^\mu\big)\cr
&=\eps^\perp_{IJ}\big(s^I\delta s^J-n^I\delta n^J\big)-\eps^\perp_{\mu\nu}\big(s^\mu\delta s^\nu-n^\mu\delta n^\nu\big)\cr
&=\sqrt{q}\big(s_I\delta n^I-n_I\delta s^I\big)-\sqrt{q}\big(s_\mu\delta n^\mu-n_\mu\delta s^\mu\big)\cr
&=2\sqrt{q}\,s_I\delta n^I-\sqrt{q}\,s_\mu\delta n^\mu\cr
&=2\btE_I\delta n^I-\sqrt{q}\,s_\mu\delta n^\mu,
\ee
where in the second to last step we have used the foliation preserving condition $\delta n_\mu\propto n_\mu$.

\subsection{Relationship between the normal, flux, and coframe}
\la{appendix:En}

Here we prove some useful relations involving the normal coframe 1-form, and in particular \eqref{E in terms of n}. First, we have
\be
\f{1}{2}(*\n)_{IK}\wedge{\n_J}^K
&=\f{1}{4}\eps_{IKML}\n^{ML}\wedge{\n_J}^K\cr
&=\f{1}{4}\eps_{IKML}(n^M\te^L-n^L\te^M)\wedge(n_J\te^K-n^K\te_J)\cr
&=\f{1}{2}\eps_{IKML}n^M\te^L\wedge(n_J\te^K-n^K\te_J)\cr
&=\f{1}{2}\eps_{IKML}n^Mn_J(\te^L\wedge\te^K)\cr
&=\f{1}{2}\teps_{ILK}n_J(\te\wedge\te)^{LK}\cr
&=\tE_In_J.
\ee
For the second equality, a direct calculation gives
\be
{\n^{[I}}_K\wedge\n^{J]K}=(n^I\te_K-n_K\te^I)\wedge(n^J\te^K-n^K\te^J)=\sigma(\te\wedge\te)^{IJ}.
\ee
Denoting by $[\cdot\,,\cdot]$ the matrix commutator \eqref{def commutator}, these two relations can be rewritten respectively in the form
\be
\tE^In^J-\tE^Jn^I=-\f{1}{2}[*\n\wedge\n]^{IJ}=-\f{1}{2}*[\n\wedge\n]^{IJ}=\f{1}{2}{\eps^{IJ}}_{KL}{\n^{[K}}_M\wedge\n^{L]M}=\f{1}{2}{\eps^{IJ}}_{KL}{\n^K}_M\wedge\n^{LM}
\ee
and
\be
(\te\wedge\te)^{IJ}=-\f{\sigma}{2}[\n\wedge\n]^{IJ}.
\ee
From this we also get
\be
\teps^{IJK}\tE_K=-\sigma(\te\wedge\te)^{IJ}=\f{1}{2}[\n\wedge\n]^{IJ}.
\ee
Using these ingredients we can decompose the wedge product of coframes as
\begin{subequations}
\be
(e\wedge e)^{IJ}&=\un\wedge\n^{IJ}+(\te\wedge\te)^{IJ}=\un\wedge\n^{IJ}-\sigma{\teps^{IJ}}_K\tE^K,\\
*(e\wedge e)^{IJ}&=\un\wedge*\n^{IJ}+2\sigma\tE^{[I}n^{J]}.
\ee
\end{subequations}
Finally, using this we can write the horizontal part of the $E^{IJ}=(*+\beta)(e\wedge e)^{IJ}$ in terms of the 1-forms $\n^{IJ}$ as
\be
\tE^{IJ}=\f{\sigma}{2}\left({(*\n)^{[I}}_K\wedge\n^{J]K}+\beta{\n^{[I}}_K\wedge\n^{J]K}\right),
\ee
or more compactly
\be
\tE^{IJ}=-\f{\sigma}{4}(*+\beta)[\n\wedge\n]^{IJ}.
\ee

\section{Relationship between ECH and GR Lagrangians}\la{AppD}

Let us consider the codimension-1 Lagrangian
\be
L_{\ECH/\GR}\coloneqq\f{1}{2}\Big(\sigma*(e\wedge e)_{IJ}\wedge\rd_\omega n^In^J-\beta e_I\wedge\rd_\omega e^I\Big).
\ee
Our goal is to show \eqref{variation ECH/GR Lagrangian}, or in other words that $L_\ECH-\rd L_{\ECH/\GR}$ is the Lagrangian $L_\GR$ written in tetrad variables. To show this we establish the relationship at the level of the symplectic potentials by varying the boundary Lagrangian. We have the variation
\be
\delta L_{\ECH/\GR}
&=\f{1}{2}\Big(2\sigma*(e\wedge e)_{IK}n_Jn^K+\beta(e\wedge e)_{IJ}\Big)\wedge\delta\omega^{IJ}\cr
&\phantom{=\ }+\delta e^I\wedge\Big(\sigma\teps_{IJK}e^J\wedge\rd_\omega n^K-\beta\rd_\omega e_I\Big)\cr
&\phantom{=\ }+\sigma\Big(2*(e\wedge e)_{IJ}\wedge\rd_\omega n^I+\rd_\omega\big(*(e\wedge e)_{IJ}\big)n^I\Big)\delta n^J\cr
&\phantom{=\ }+\rd\left(\sigma*(e\wedge e)_{IJ}\delta n^In^J+\f{\beta}{2}e_I\wedge\delta e^I\right).
\ee
Going on-shell using $\rd_\omega e^I\simeq0$, denoting $K^I=\rd_\omega n^I$, pulling-back to $\Sigma$, and using the definitions $\tE_I=*(\te\wedge\te)_{IJ}n^J$ and $\tP^I=-\sigma(\tK\times\te)^I$, we get that the third line vanishes. Indeed, the last term vanishes on-shell, and the first term vanishes by virtue of \eqref{vanishing EK delta n}. Noticing that the second line on-shell and on $\Sigma$ is
\be
\sigma\teps_{IJK}\delta\te^I\wedge\te^J\wedge\tK^K=-\tP_I\wedge\delta\te^I=-\theta_\GR,
\ee
we are therefore left with the identity
\be
\delta L_{\ECH/\GR}
&\stackrel{S}{\simeq}\theta_\ECH-\theta_\GR-\rd\theta^S_\ECH\cr
&=\theta_\ECH-\theta_\GR-\rd\theta_{\ECH/\GR}.
\ee
This is the exact analog of the formula for $\delta L_{\EH/\GR}$ which was derived in \cite{Freidel:2020xyx}.

\section{Alternative decomposition of the ECH potential}
\la{appendix:AE potential}

Here we present an alternative way to write the ECH symplectic potential. It is a Lorentz covariant version of the usual LQG potential in connection and triad variables. At the difference with what we have done in the main text, here the contribution of the topological Holst term is put in the bulk and not on the boundary. This leads to the covariant (i.e. without fixing $n^I$ in the time gauge) parametrization of the bulk degrees of freedom with the Ashtekar--Barbero connection $A^I$ conjugated to $E_I$. We did not use this decomposition because we have chosen instead to write the bulk piece as the GR potential, and to push the dependency on the Barbero--Immirzi parameter on the boundary.

Using \eqref{E on Sigma}, one can write
\be\la{covariant AB connection potential}
{\sigma}\Theta_\ECH
&=\f{\sigma}{2}\int_\Sigma E_{IJ}\wedge\delta\omega^{IJ}\cr
&=\int_\Sigma\tE_I\wedge\delta(\omega-\beta*\omega)^{IJ}n_J\cr
&=\int_\Sigma\left(\tE_I\wedge\delta(\CA^{IJ}n_J)-\tE_I\wedge\CA^{IJ}\delta n_J\right)\cr
&=\int_\Sigma\left(\tE_I\wedge\delta(\rd n^I+\CA^{IJ}n_J)-\tE_I\wedge\delta\rd n^I-\tE_I\wedge\CA^{IJ}\delta n_J\right)\cr
&=\int_\Sigma\left(\tE_I\wedge\delta A^I+\pi_I\delta n^I\right)-\int_S\tE_I\delta n^I,
\ee
where we have introduced
\be
\CA^{IJ}\coloneqq(1-\beta*)\omega^{IJ},\q A^I\coloneqq\rd_\CA n^I,\q\pi_I\coloneqq\rd_\CA\tE_I.
\ee
This shows that the gravitational ECH phase space can be parametrized in terms of the two bulk canonical pairs $(\tE_I,A^I)$ and $(n^I,\pi_I)$, and the corner pair $(\tE_I,n^I)$. In the bulk, the first pair represents $9$ degrees of freedom (since we have $\tE_In^I=0$ and $A^In_I=0$) and the second pair $3$, for a total of 12. This is compensated by $6$ Lorentz transformations and $4$ diffeomorphisms, so a total of $10$ constraints, which leaves indeed 2 degrees of freedom. In the time gauge $n^I=(1,0,0,0)$ the momentum $A^I$ becomes the Ashtekar--Barbero connection $A^i\coloneqq\CA^{i0}$.

For the sake of curiosity, we can match the expression \eqref{covariant AB connection potential} with the various ingredients of the bulk-corner decomposition which we have performed in Section \ref{sec:3+1}. In order to do so, let us first notice that \eqref{tensor decomposition} implies that
\be
\tE_I\wedge(*\omega)^{IJ}\delta n_J=\sigma(\omega_\per\times\tE)^I\delta n_I.
\ee
Using this and the fact that $(*\omega)^{IJ}n_J=\Gamma_\para^I$, we can write the Holst piece of the ECH potential as
\be
\f{1}{2}\int_\Sigma(\te\wedge\te)_{IJ}\wedge\delta\omega^{IJ}
&=\int_\Sigma\tE_I\wedge\delta(*\omega)^{IJ}n_J\cr
&=\int_\Sigma\left(\tE_I\wedge\delta\big((*\omega)^{IJ}n_J\big)-\tE_I\wedge(*\omega)^{IJ}\delta n_J\right)\cr
&=\int_\Sigma\left(\tE_I\wedge\delta\Gamma^I_\para-\sigma(\omega_\per\times\tE)^I\delta n_I\right).
\ee
With this, the full ECH potential can be written in the form
\be
\Theta_\ECH=\int_\Sigma\left(\tE_I\wedge\delta(\sigma\tK^I+\beta\Gamma^I_\para)+\sigma\big(\tilde{\rd}_\Gamma\tE_I-\beta(\omega_\per\times\tE)^I\big)\delta n^I\right)-\sigma\int_S\tE_I\delta n^I,
\ee
so that even on-shell of the Gauss constraint there is a pair in the bulk involving the normal $n^I$. One can then check that the momenta conjugated to $\tE_I$ and $n^I$ are indeed $A^I$ and $\pi_I$ introduced in \eqref{covariant AB connection potential}.

\section{Relative diffeomorphism charge}
\la{appendix:relative diffeo}

\subsection{From the difference of charges}
\la{appendix:relative diffeo 1}

Here we establish \eqref{relative diffeo charge} by a direct subtraction of the ECH and GR charges. It is actually possible to start this calculation by decomposing the BF diffeomorphism charge \cite{Mondragon:2004nw}.
\be
\CH^S_\BF[\xi]=\f{1}{2}\int_S\xi\ip\omega_{IJ}B^{IJ}.
\ee
This decomposition follows almost verbatim the decomposition of the potential \eqref{bare BF potential}. We first have
\be
\CH^S_\BF[\xi]=\int_S\left(\sigma\tB_I\xi\ip\omega^{IJ}n_J+\f{1}{2}\ts_I\wedge(\xi\ip\omega^{IJ})\ts_J\right).
\ee
We evaluate the two terms separately. For the first term, we use \eqref{decomposition of omega} to get
\be
\xi\ip \omega^{IJ}n_J=\xi\ip K^I+\xi\ip\Gamma^{IJ}n_J,
\ee
and \eqref{covariant Gamma of n} to write
\be
\xi\ip\rd_\Gamma n^I=\xi\ip\rd n^I+\xi\ip\Gamma^{IJ}n_J=\CL_\xi n^I+\xi\ip\Gamma^{IJ}n_J=0.
\ee
Together, this gives 
\be
\xi\ip  \omega^{IJ}n_J=\xi\ip K^I -\CL_\xi n^I.
\ee
We this we can therefore write
\be
B_I\xi\ip\omega^{IJ}n_J=B_I(\xi\ip K^I -\CL_\xi n^I)=\xi\ip(B_I\wedge K^I)-(\xi\ip B_I)K^I-B_I\CL_\xi n^I.
\ee
For the second term we have
\be
\ts_I\wedge(\xi\ip\omega^{IJ})\ts_J
&=\ts_I\wedge(\xi\ip\Gamma^{IJ})\ts_J\cr
&=\ts_I\wedge\big(\xi\ip(\Gamma^{IJ}\wedge\ts_J)+\Gamma^{IJ}(\xi\ip\ts_J)\big)\cr 
&=\ts_I\wedge\big(\xi\ip(\rd_\Gamma\ts^I-\rd\ts^I )+\rd_\Gamma(\xi\ip\ts^I)-\rd(\xi\ip\ts^I)\big)\cr 
&=\ts_I \wedge\big(\xi\ip\rd_\Gamma\ts^I+\rd_\Gamma(\xi\ip \ts^I)-\CL_\xi \ts^I\big)\cr
&=\rd_\Gamma\ts^I(\xi\ip\ts_I)-(\xi\ip\rd_\Gamma\ts^I)\wedge\ts_I-\rd\big(\ts_I(\xi\ip\ts^I)\big)-\ts_I\wedge\CL_\xi\ts^I\cr
&=2\rd_\Gamma\ts_I(\xi\ip\ts^I)-\xi\ip(\ts_I\wedge\rd_\Gamma\ts^I)-\rd\big(\ts_I(\xi\ip\ts^I)\big)-\ts_I\wedge\CL_\xi\ts^I.
\ee
One can note the ressemblance with \eqref{second term in BF pot decomp} with $\delta$ there traded for $\xi\ip$ here. Putting this together, and focusing on the case of tangent diffeomorphisms, we get that the BF diffeomorphism charge decomposes in the form
\be
\CH^S_\BF[\xi]=-\int_S\left(\sigma\xi\ip\tB_I\tK^I+\sigma\tB_I\CL_\xi n^I-\tilde{\rd}_\Gamma\ts_I(\xi\ip\ts^I)+\f{1}{2}\ts_I\wedge\CL_\xi\ts^I\right).
\ee
Imposing the simplicity constraints \eqref{simplicity 2} leads to the ECH diffeomorphism charge
\be
\CH^S_\ECH[\xi]=-\int_S\left(\sigma\xi\ip\tE_I\tK^I+\sigma\tE_I\CL_\xi n^I-\beta\tilde{\rd}_\Gamma\te_I(\xi\ip\te^I)+\f{\beta}{2}\te_I\wedge\CL_\xi\te^I\right).
\ee
Going on-shell and using the definition of $\tE^I$ together with the momentum $\tP^I=-\sigma(\tK\times\te)^I$, we finally get
\be
\CH^S_\ECH[\xi]\simeq\int_S\xi\ip\te^I\tP_I-\int_S\left(\sigma\tE_I\CL_\xi n^I+\f{\beta}{2}\te_I\wedge\CL_\xi\te^I\right).
\ee
For $\sigma=-1$, the relative charge is therefore
\be\la{H-ECH/GR}
\CH^S_{\ECH/\GR}[\xi]=\CH^S_\ECH[\xi]-\CH^S_\GR[\xi]=\int_S\left(\tE_I\CL_\xi n^I-\f{\beta}{2}\te_I\wedge\CL_\xi\te^I\right).
\ee

\subsection{From the relative symplectic structure}
\la{appendix:relative diffeo 2}

Let us rewrite the relation \eqref{ECH GR Omegas} between the GR and ECH symplectic structures using the 1-forms \eqref{nmode}. As shown in Section \ref{sec:boundary with nIJ}, we can write
\be
\Omega_\text{GR}=\Omega_\ECH-\Omega_{\ECH/\GR}=\Omega_\ECH+\f{\sigma}{4}\int_S(*+\beta)\delta\n_{IJ}\wedge\delta\n^{IJ}.
\ee
We now want to use this relationship between the symplectic structures in order to relate the GR, ECH, and relative charges.

Let us first notice that the contraction of a diffeomorphism with the boundary term is integrable by virtue of the identity
\be
-\CL_\xi\ipp\left(\int_S\delta\n_{IJ}\wedge\delta\n^{IJ}\right)
&=2\int_S\delta\n_{IJ}\wedge\CL_\xi\n^{IJ}\cr
&=2\int_S\delta\n_{IJ}\wedge\big(\rd(\xi\ip\n^{IJ})+\xi\ip(\rd\n^{IJ})\big)\cr
&=2\int_S\delta\n_{IJ}\wedge\rd(\xi\ip\n^{IJ})+\xi\ip\delta\n_{IJ}\rd\n^{IJ}\cr
&=2\int_S\delta\n_{IJ}\wedge\rd(\xi\ip\n^{IJ})+\n^{IJ}\wedge\rd(\xi\ip\delta\n_{IJ})\cr
&=2\int_S\delta\big(\n_{IJ}\wedge\rd(\xi\ip\n^{IJ})\big).
\ee
This tells us that the relative charge is
\be
\CH^S_{\ECH/\GR}[\xi]=-\f{\sigma}{2}\int_S(*+\beta)\n_{IJ}\wedge\CL_\xi\n^{IJ}.
\ee
We can then contract a diffeomorphism with the above symplectic structures to find the following relation between the corner charges:
\be\la{BY charge in terms of n}
\CH^S_\text{GR}[\xi]
&=\CH^S_\ECH[\xi]-\CH^S_{\ECH/\GR}[\xi]\cr
&=\int_S\xi\ip\omega_{IJ}\tE^{IJ}+\f{\sigma}{2}\int_S(*+\beta)\n_{IJ}\wedge\rd(\xi\ip\n^{IJ})\cr
&=\int_S\omega_{IJ}\wedge\xi\ip\tE^{IJ}+\f{\sigma}{2}\int_S(*+\beta)\n_{IJ}\wedge\rd(\xi\ip\n^{IJ})\cr
&=\sigma\int_S\omega_{IJ}\wedge(*+\beta)(\xi\ip{\n^{[I}}_K)\n^{J]K}+\f{\sigma}{2}\int_S(*+\beta)\n_{IJ}\wedge\rd(\xi\ip\n^{IJ})\cr
&=\f{\sigma}{2}\int_S(*+\beta)\n_{IJ}\wedge\rd_\omega(\xi\ip\n^{IJ}),
\ee
where for the third equality we have used the fact that $\xi$ is tangent to $S$, and for the fourth one the simplicity constraint \eqref{E in terms of n}. Our goal is to show, as a consistency check, that this charge written on the right-hand side in terms of $\n^{IJ}$ is indeed the Brown--York charge $\CH^S_\text{GR}[\xi]$.

For this, let us study the two pieces of this expression separately. The second term, proportional to the Barbero--Immirzi parameter, can be rewritten in metric form as
\be
\f{\sigma}{2}\int_S\n_{IJ}\wedge\rd_\omega(\xi\ip\n^{IJ})
&=\int_S\n_{IJ}\wedge\rd_\omega(n^I\xi^J)\cr
&=\int_S\n_{IJ}\wedge(\tK^I\xi^J+n^I\rd_\omega\xi^J)\cr
&=\sigma\int_S(n_I\te_J-n_J\te_I)\wedge(\tK^I\xi^J+n^I\rd_\omega\xi^J)\cr
&\simeq\int_S\te_I\wedge\rd_\omega\xi^I\cr
&\simeq\int_S\eps^{\mu\nu\rho\sigma}n_\mu s_\nu\te_{I\rho}\nabla_\sigma(\xi^\alpha\te^I_\alpha)\cr
&=-\int_S\eps^{\mu\nu\rho\sigma}n_\mu s_\nu\nabla_\rho\xi_\sigma,
\ee
where we have used \eqref{horizontal torsion} and introduced $\xi^I=\xi\ip\te^I$. This is the trivial ``topological'' Komar charge as in \cite{DePaoli:2018erh}. The first term in \eqref{BY charge in terms of n}, however, can be rewritten in terms of the momentum $\tP^I$ as
\be
\f{\sigma}{2}\int_S(*\n)_{IJ}\wedge\rd_\omega(\xi\ip\n^{IJ})
&=\int_S(*\n)_{IJ}\wedge(K^I\xi^J+n^I\rd_\omega\xi^J)\cr
&=\f{\sigma}{2}\int_S\eps_{IJKL}(n^K\te^L-n^L\te^K)\wedge(K^I\xi^J+n^I\rd_\omega\xi^J)\cr
&=\f{\sigma}{2}\int_S\eps_{IJKL}(n^K\te^L-n^L\te^K)\wedge K^I\xi^J\cr
&=-\sigma\int_S\eps_{IJKL}\xi^IK^J\wedge\te^Kn^L\cr
&=\int_S\xi^I\tP_I,
\ee
which as expected is the Brown--York charge.

\section{First class nature of the gluing condition}
\la{B gluing}

In this final appendix we prove the statement that the rotated Lie algebra elements $(\vphi\JJ\vphi^{-1})^{IJ}$ commute with $\JJ^{IJ}$ and form a Lorentz algebra $\sll(2,\mathbb{C})$ with orientation opposite to $\JJ^{IJ}$.  This explains why the matching condition $B^{IJ}\stackrel{S}\simeq(\vphi\JJ\vphi^{-1})^{IJ}$ at the corner is first class.

In order to get this result, we use the expression \eqref{cJJ} for the corner generator and the definition $\bd=\delta-\varphi^{-1}\delta\varphi$ to rewrite the edge mode contribution to the corner potential \eqref{edge mode potential} as
\be
-\bTh_\BF^S=\int_S\left(\sigma\BB_I\delta\n^I+\f{\beta}{2}\e_I\wedge\delta\e^I-\f{1}{2}(\varphi^{-1}\delta\varphi)^{IJ}\JJ_{IJ}\right).
\ee
From this we can read the commutators of $\JJ[\alpha]=\f{1}{2}\int_S\alpha^{IJ}\JJ_{IJ}$ and $\varphi$, which are
\begin{subequations}
\be
\{\JJ[\alpha],\JJ[\beta]\}&=\JJ\big[[\alpha,\beta]\big],\\
\{\JJ[\alpha],\varphi(x) \}&=\varphi(x)\alpha,\\
\{\varphi(x),\varphi(y)\}&=0.
\ee
\end{subequations}
Given these commutation relations, we see that the generator $\JJ$ commutes with the corner generator of gauge transformations, i.e.
\be
\{\JJ[\alpha],\JJ[\varphi^{-1}\beta\varphi]\}=0.
\ee
This commutation relation follows from the fact that $\JJ[\alpha]$ acts by right multiplication on $\varphi$ while $\JJ[\varphi^{-1}\beta\varphi ]$ acts instead by left multiplication. The commutator of the dressed charges is then given by
\be
\{\JJ[\varphi^{-1}\alpha\varphi],\JJ[\varphi^{-1}\beta\varphi]\}=-\JJ\big[\varphi^{-1}[\alpha,\beta]\varphi\big].
\ee
This shows that if we have generators $B^{IJ}$satisfying a local Lorentz algebra $\{B[\alpha],B[\beta]\}=B\big[[\alpha,\beta]\big]$, the constraint
$B[\alpha]-\JJ[\alpha]=0$ is second class while $B[\alpha]-\JJ[\varphi^{-1}\alpha\varphi]=0$ is first class, as announced.

\bibliographystyle{bib-style2}
\bibliography{Biblio}

\end{document}